\documentclass[twoside, twocolumn]{article}     

\usepackage[T1]{fontenc}
\usepackage{times}

\usepackage[pdftex]{graphics}
\usepackage{amsmath}
\usepackage{amssymb}
\usepackage{deluxetable}
\usepackage{natbib}
\bibpunct[, ]{(}{)}{;}{a}{}{,}

\graphicspath{{figures/}}

\usepackage[preprint]{amsstyle}
\newcommand  {\massavg}[1]{\ensuremath{\overline{#1}^{*}}}
\newcommand  {\about} {\mathop{\sim}\!}
\newcommand  {\divh}  {\vec{\nabla}_h \cdot}
\newcommand  {\divy}  {\mathop{\mathrm{div}}}
\renewcommand{\div}   {\vec{\nabla} \cdot}
\newcommand  {\grad}  {\vec{\nabla}}
\newcommand  {\gradh} {\vec{\nabla}_h}

\reference{\textit{Journal of the Atmospheric Sciences} (in press)}

\begin{document}


\title{Mechanisms of jet formation on the giant planets}

\author{Junjun Liu  and Tapio Schneider
  \thanks{\emph{Corresponding author address:}
    Tapio Schneider, California Institute of Technology, Mail Code
    100-23, 1200 E.\ California Blvd., Pasadena, CA 91125.
    E-mail: tapio@caltech.edu}%
    }%
\address{California Institute of Technology, Pasadena, California}

\AbstractText{The giant planet atmospheres exhibit alternating
  prograde (eastward) and retrograde (westward) jets of different
  speeds and widths, with an equatorial jet that is prograde on
  Jupiter and Saturn and retrograde on Uranus and Neptune. The jets
  are variously thought to be driven by differential radiative heating
  of the upper atmosphere or by intrinsic heat fluxes emanating from
  the deep interior.  But existing models cannot account for the
  different flow configurations on the giant planets in an
  energetically consistent manner. Here a three-dimensional general
  circulation model is used to show that the different flow
  configurations can be reproduced by mechanisms universal across the
  giant planets if differences in their radiative heating and
  intrinsic heat fluxes are taken into account. Whether the equatorial
  jet is prograde or retrograde depends on whether the deep intrinsic
  heat fluxes are strong enough that convection penetrates into the
  upper troposphere and generates strong equatorial Rossby waves
  there. Prograde equatorial jets result if convective Rossby wave
  generation is strong and low-latitude angular momentum flux
  divergence owing to baroclinic eddies generated off the equator is
  sufficiently weak (Jupiter and Saturn). Retrograde equatorial jets
  result if either convective Rossby wave generation is weak or absent
  (Uranus) or low-latitude angular momentum flux divergence owing to
  baroclinic eddies is sufficiently strong (Neptune). The different
  speeds and widths of the off-equatorial jets depend, among other
  factors, on the differential radiative heating of the atmosphere and
  the altitude of the jets, which are vertically sheared. The
  simulations have closed energy and angular momentum balances that
  are consistent with observations of the giant planets. They exhibit
  temperature structures closely resembling those observed, and make
  predictions about as-yet unobserved aspects of flow and temperature
  structures.}

\maketitle

\section{Introduction}

\begin{figure*}[!htb]
\centering \includegraphics{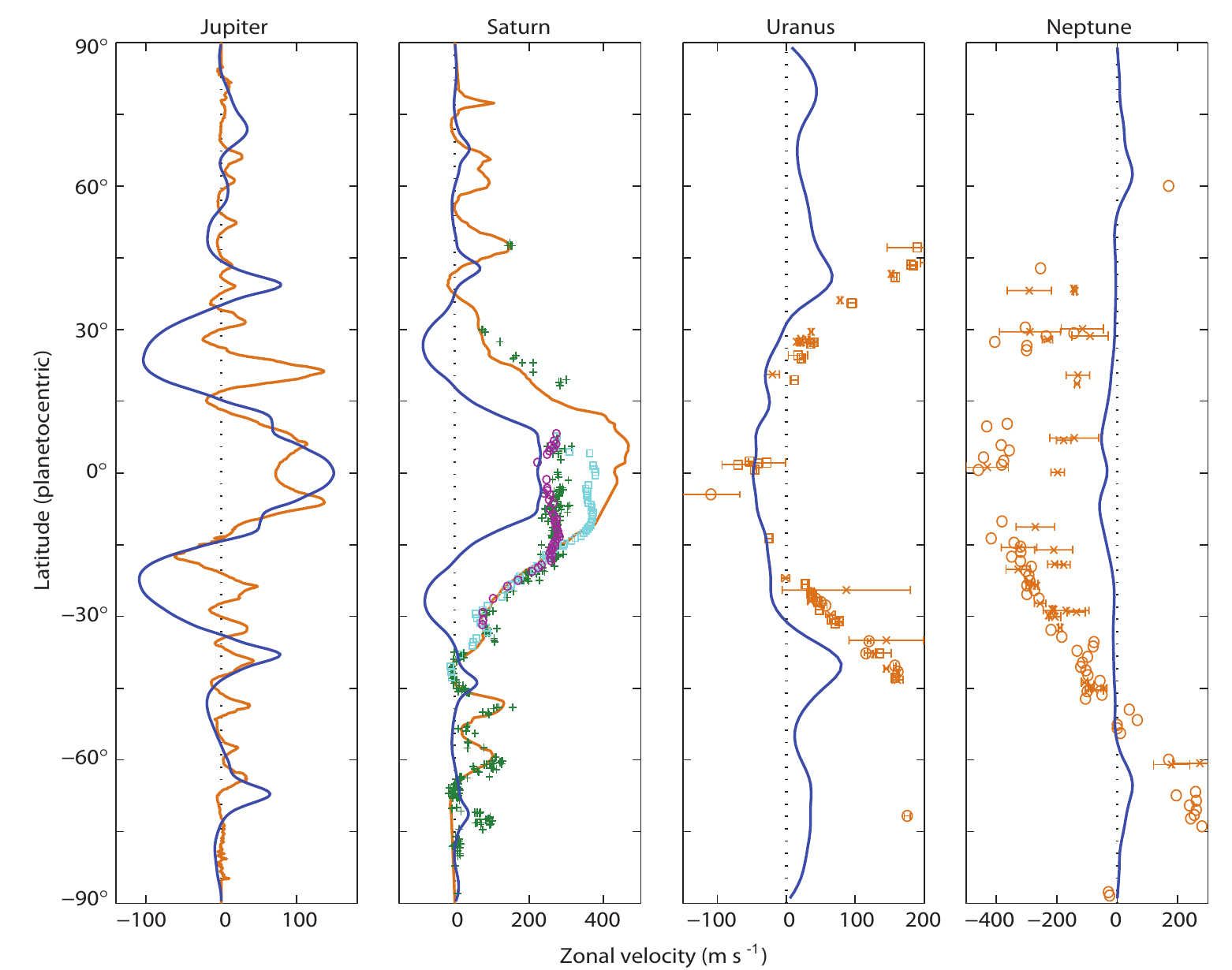}
\caption{Mean zonal velocities in the upper atmosphere of the giant
  planets from observations and simulations.  Jupiter: observations
  from the \emph{Cassini} spacecraft \citep{Porco03} (orange line),
  and in simulation at $0.75$~bar (dark blue line).  Saturn:
  observations from the \emph{Voyager} spacecraft (orange line), from
  the \emph{Hubble Space Telescope} (HST) \citep{Sanchez-Lavega03}
  (green crosses), from the \emph{Cassini} spacecraft at $\about
  0.06$~bar (magenta circles) and at $\about 0.7$~bar (light blue
  squares) \citep{Sanchez-Lavega07}, and in simulation at $0.1$~bar
  (dark blue line). Uranus: observations from the \emph{Voyager}
  spacecraft (orange circles), HST (orange crosses) \citep{Hammel01},
  the Keck telescope (orange squares) \citep{Hammel05}, and in
  simulation at $25.0$~mbar (dark blue line). Neptune: observations
  from the \emph{Voyager} spacecraft (orange circles) and from HST
  (orange crosses) \citep{Sromovsky01}, and in simulation at
  $25.0$~mbar (dark blue line). Differences between the statistically
  identical northern and southern hemispheres in the simulations are
  indicative of the sampling variability of the averages.}
\label{fig_flow_lat}
\end{figure*}

Among the most striking features of the giant planets are the
alternating zonal jets. As shown in Fig.~\ref{fig_flow_lat}, Jupiter
and Saturn have prograde equatorial jets (superrotation) that peak at
$\about 100\,\mathrm{m \, s^{-1}}$ and $\about 200$--$400\,\mathrm{m
  \, s^{-1}}$, depending on the vertical level considered. Uranus and
Neptune have retrograde equatorial jets (subrotation) that peak at
$\about 100\,\mathrm{m \, s^{-1}}$ and $\about 150$--$400\,\mathrm{m
  \, s^{-1}}$.  Jupiter and Saturn have multiple off-equatorial jets
in each hemisphere; Uranus and Neptune have only a single
off-equatorial jet in each hemisphere.  Despite decades of study with
a variety of flow models, it has remained obscure how these different
flow configurations come about \citep{Vasavada05}.

Existing models posit as the driver of the flow either the
differential radiative heating of the upper atmosphere
\cite[e.g.,][]{Williams79,Williams03a} or the intrinsic heat fluxes
emanating from the deep interior
\citep[e.g.,][]{Busse76,Heimpel05,Aurnou07,Chan08,Kaspi09}.  However,
none of these models can account for the existence of equatorial
superrotation on Jupiter and Saturn and equatorial subrotation on
Uranus and Neptune with radiative heating, intrinsic heat fluxes, and
other physical parameters consistent with observations.

For example, deep-flow models that posit intrinsic heat fluxes as the
sole driver of the flow can generate equatorial superrotation, but
they use heat fluxes more than $10^6$ times larger than those observed
\citep[e.g.,][]{Heimpel07}. They generate equatorial subrotation only
with intrinsic heat fluxes even stronger than those for which they
generate superrotation \citep{Aurnou07}, although the intrinsic heat
fluxes on the subrotating planets (Uranus and Neptune) are weaker than
those on the superrotating planets (Jupiter and Saturn).  The
relevance of such deep-flow models is further called into question by
the eddy angular momentum fluxes they imply. Their meridional eddy
fluxes of angular momentum per unit volume (taking density variations
into account) have a barotropic structure: they extend roughly along
cylinders concentric with the planet's spin axis over the entire depth
of the fluid, typically to pressures of order $10^6$~bar
\citep[e.g.,][their Fig.~10]{Kaspi09}. But the eddy angular momentum
fluxes inferred from tracking cloud features in Jupiter's and Saturn's
upper tropospheres indicate that the mean conversion rate from eddy to
mean-flow kinetic energy is of order $10^{-5} \, \mathrm{W \ m^{-3}}$
\citep{Ingersoll81,Salyk06,DelGenio07}. If the observed
upper-tropospheric eddy fluxes of angular momentum per unit volume
extended unabatedly over a layer of 50~km thickness (e.g., from about
0.3 to 2.5~bar pressure on Jupiter, or from about 0.3 to 0.9~bar
pressure on Saturn), and if vertical zonal flow variations over this
layer are not dramatic, the total energy conversion rate would be
about $0.5\,\mathrm{W \, m^{-2}}$. This is already $\about 4\%$ of the
total energy uptake of the atmosphere from intrinsic heat fluxes and
absorption of solar radiation for Jupiter, or $\about 11\%$ for
Saturn. But the limited thermodynamic efficiency of atmospheres allows
only a fraction of the total atmospheric energy uptake to be used to
generate eddy kinetic energy \citep{Lorenz55,Peixoto92}. The
observations of Jupiter and Saturn therefore imply that eddy angular
momentum fluxes cannot extend unabatedly over great depths and must
have a baroclinic structure. Barotropic eddy angular momentum fluxes
that extend to depths of order $10^6$~bar, with upper-atmospheric
fluxes of similar scale and magnitude as those observed, are only
possible in deep-flow models if the driving heat fluxes are several
orders of magnitude greater than observed.

Similarly, shallow-flow models that posit differential radiative
heating as the driver of the flow can generate equatorial
superrotation, but they require artifices such as additional
equatorial heat or wave sources that have no clear physical
interpretation \citep[e.g.,][]{Williams03a, Williams03b, Yamazaki05,
  Lian08}. It is unclear in those and other shallow-flow models
\citep[e.g.,][]{Scott08} what physical characteristics distinguish the
superrotating planets from the subrotating
planets.\footnote{\citet{Lian10} claim that different rates of latent
  heat release in phase changes of water may be responsible for
  superrotation on Jupiter and Saturn and subrotation on Uranus and
  Neptune. However, they impose latent heat fluxes at the lower
  boundary of their model that are not consistent with the observed
  energetics of the planets. Similar to the simulations of
  \citet{Aurnou07}, they require stronger energy (latent heat) fluxes
  to generate subrotation than to generate superrotation. For example,
  the latent heat fluxes are of order $10$--$20\,\mathrm{W\,m^{-2}}$
  in their Jupiter and Saturn simulations and of order
  $1500\,\mathrm{W\,m^{-2}}$ in their Uranus/Neptune simulation
  (Y.~Lian, pers.\ communication, 2010). The latter are several orders
  of magnitude larger than the observed intrinsic heat fluxes or
  absorbed radiative fluxes (Table~1), which would have to drive any
  latent heat fluxes (energy would be required to evaporate the
  condensate that falls from the upper atmosphere into deeper
  layers).}

In \citet{Schneider09} (SL09 hereafter), we postulated that prograde
equatorial jets on the giant planets occur when intrinsic heat fluxes
are strong enough that Rossby waves generated convectively in the
equatorial region transport angular momentum toward the
equator. Multiple off-equatorial jets, by contrast, form as a result
of baroclinic instability owing to the differential radiative heating
of the upper atmosphere. We introduced a general circulation model
(GCM) and demonstrated with it that the postulated mechanisms can
account qualitatively for large-scale flow structures observed on
Jupiter. Here we use simulations with essentially the same GCM, with
closed energy and angular momentum balances that are consistent with
observations, to demonstrate universal formation mechanisms of jets on
all the giant planets.  We show that the different flow configurations
on the giant planets can be explained through consideration of the
different roles played by intrinsic heat fluxes and solar radiation in
generating atmospheric waves and instabilities.

Section~\ref{s:GCM} briefly describes the GCM.
Section~\ref{s:Simulations} shows simulation results for Jupiter,
Saturn, Uranus, and Neptune. Section~\ref{s:Mechanisms} discusses the
formation mechanisms of the jets in the simulations and confirms the
postulated mechanisms through control simulations.
Section~\ref{s:AM_balance} discusses what the upper-atmospheric fluid
dynamics, on which we focus, imply about flows at greater depth on the
giant planets. Section~\ref{s:Conclusion} summarizes the conclusions
and their relevance for available and possible future observations.

\section{General circulation model}
\label{s:GCM}

With current computational resources, it is not feasible to simulate
flows deep in giant planet atmospheres, where radiative relaxation
times are measured in centuries and millennia, while at the same time
resolving the energy-containing eddies in the upper
atmospheres. Therefore, we focus on flows in the upper atmospheres,
using a GCM that solves the hydrostatic primitive equations for a dry
ideal-gas atmosphere in a thin spherical shell. The model is
essentially that introduced for Jupiter in SL09, but here we use it to
also simulate Saturn, Uranus, and Neptune.\footnote{The Jupiter
  simulation here differs slightly from that in SL09 in that poorly
  constrained drag parameters in it are chosen to be the same as in
  the simulations of the other giant planets presented here.}
Parameters such as the planetary rotation rate, gravitational
acceleration, and material properties of the atmosphere in each
simulation are those of the planet being simulated. The resolution in
each simulation (T85 to T213 spectral resolution in the horizontal and
30 or 40 levels in the vertical) is sufficient to resolve baroclinic
instability and the energy-containing eddies in the upper
atmosphere. The GCM and the simulations are described in detail in the
appendix, and Table~1 lists the parameters; here we only give a brief
overview.
\begin{table*}[!htb]
  \centering
  \caption{Simulation Parameters}\label{table_all_giants}
  \begin{tabular}{lllll}
    \hline\hline\\[-2ex]
    \multicolumn{1}{c}{Parameter, symbol} & \multicolumn{1}{c}{Jupiter}
    & \multicolumn{1}{c}{Saturn} & \multicolumn{1}{c}{Uranus}
    & \multicolumn{1}{c}{Neptune} \\[.25ex]
    \hline\\[-2ex]
     Planetary radius, $a$ $(10^{6} \, \mathrm{m})$ &
    $69.86 $\tablenotemark{a} &
    $57.32 $\tablenotemark{a} &
    $25.27 $\tablenotemark{b} &
    $24.55 $\tablenotemark{b}
     \\

    Planetary angular velocity, $\Omega$ $(10^{-4} \, \mathrm{s})$ &
    $1.7587 $\tablenotemark{b} &
    $1.6388 $\tablenotemark{b} &
    $1.0124 $\tablenotemark{b} &
    $1.0834 $\tablenotemark{b}
    \\

    Gravitational acceleration, $g$ $(\mathrm{m \, s^{-2}})$&
    $26.0$\tablenotemark{b} &
    $10.55$\tablenotemark{b} &
    $8.94$\tablenotemark{b} &
    $11.2$\tablenotemark{b}
    \\

    Specific gas constant, $R$ $(\mathrm{J \, kg^{-1} \, K^{-1}})$ &
    $3605.38 $\tablenotemark{b} &
    $4016.4  $\tablenotemark{b} &
    $3149.2  $\tablenotemark{b} &
    $3197.7  $\tablenotemark{b}
     \\

    Adiabatic exponent, $\kappa$ &
    $2/7$ &
    $2/7$ &
    $2/7$ &
    $2/7$ \\

    Specific heat capacity, $c_p=R/\kappa$ $(10^{4} \, \mathrm{J \, kg^{-1} \, K^{-1}})$ &
    $1.26$ &
    $1.41$ &
    $1.10$ &
    $1.12$
    \\

    Solar constant $F_0$ $(\mathrm{W\, m^{-2}})$  &
    $50.7 $\tablenotemark{c} &
    $14.9 $\tablenotemark{c} &
    $3.71 $\tablenotemark{c}&
    $1.52 $\tablenotemark{c}
     \\

    Intrinsic heat flux $(\mathrm{W\, m^{-2}})$ &
    $5.7 $\tablenotemark{d} &
    $2.01$\tablenotemark{e} &
    $0.042$\tablenotemark{e} &
    $0.433$\tablenotemark{e}
     \\

    Bond albedo, $r_\infty$ &
    $0.343$\tablenotemark{f}  &
    $0.342$\tablenotemark{g} &
    $0.30$\tablenotemark{b}  &
    $0.29$\tablenotemark{b}
    \\

    Single-scattering albedo, $\tilde{\omega}$ &
    $0.8$ &
    $0.8$ &
    $0.8$ &
    $0.8$
    \\

    Solar optical depth at 3~bar, $\tau_{s0}$ &
    $3.0$ &
    $3.0$ &
    $3.0$ &
    $3.0$
    \\

    Thermal optical depth at 3~bar, $\tau_{l0}$ &
    $80.0$ &
    $120.0$ &
    $60.0$ &
    $40.0$
    \\

    Drag coefficient, $k_0$ $(\mathrm{days}^{-1}=(86400\,\mathrm{s}^{-1}))$ &
    $1/100$ &
    $1/100$ &
    $1/100$ &
    $1/100$
    \\

    No-drag latitude, $\phi_0$ &
    $33^\circ$ &
    $33^\circ$ &
    $33^\circ$ &
    $33^\circ$
    \\

    Horizontal spectral resolution &
    T213 &
    T213 &
    T85 &
    T85
    \\

    Vertical levels &
    30 &
    30 &
    40 &
    40
    \\

    Cut-off wavenumber for subgrid-scale dissipation &
    100 &
    100 &
    40 &
    40\\
    \hline
    \multicolumn{5}{l}{\footnotesize{%
      ${}^a${\cite{Guillot99}};
      ${}^b${\cite{Lodders98}};
      ${}^c${\cite{Levine77}};
      ${}^d${\cite{Gierasch00}};}}\\
    \multicolumn{5}{l}{\footnotesize{%
      ${}^e${\cite{Guillot05}};
      ${}^f${\cite{Hanel81}};
      ${}^g${\cite{Hanel83}}}}
 \end{tabular}
\end{table*}

The GCM domain is a thin but three-dimensional spherical shell, which
extends from the top of the atmosphere to an artificial lower
boundary.  The mean pressure at the lower boundary is $3$~bar in all
our simulations, to minimize differences in arbitrary parameters among
them. Insolation is imposed as perpetual equinox with no diurnal cycle
at the top of the atmosphere. Absorption and scattering of solar
radiation and absorption and emission of thermal radiation are
represented in an idealized way that is consistent with observations
where they are available (primarily for Jupiter). Where radiative and
other parameters are not well constrained by observations or by
knowledge of physical properties of the planets, we set them to be
equal to the parameters for Jupiter, again to minimize differences in
unconstrained parameters among the simulations. A dry convection
scheme relaxes temperature profiles in statically unstable layers
toward a convective profile with dry-adiabatic lapse rate, without
transporting momentum in the vertical (see the appendix for details
and for a discussion of this idealization).

At the lower boundary of the GCM, a temporally constant and spatially
uniform intrinsic heat flux is imposed, with magnitude equal to the
observed intrinsic heat fluxes ($5.70$, $2.01$, and $0.04$, and
$0.43\,\mathrm{W\,m^{-2}}$ for Jupiter, Saturn, Uranus, and
Neptune). (The heat flux in the Uranus simulation corresponds to an
observational upper bound.)  Linear (Rayleigh) drag retards the flow
away from but not near the equator---a thin-shell representation of a
magnetohydrodynamic (MHD) drag that acts at great depth (at pressures
$\gtrsim 10^5$~bar), where the atmosphere becomes electrically
conducting \citep{Liu08}. Drag in a deep atmosphere affects the
angular momentum balance averaged over cylinders concentric with the
planet's spin axis, so there is no effective drag on the flow in the
upper atmosphere near the equator, in the region in which the
cylinders do not intersect the layer of MHD drag at depth
(SL09). Absent detailed knowledge of where and how the MHD drag acts
and to rule out that differences among the simulations are caused by
differences in the drag formulation, we chose the equatorial no-drag
region to extend to $\phi_0=33^{\circ}$ latitude and the drag
coefficient outside this region to be the same in all
simulations. Section~\ref{s:Mechanisms}\ref{s:drag} discusses the
effect of this drag formulation on our simulation results, and
section~\ref{s:AM_balance} and the appendix provide further
justification for it.

We show simulation results from statistically steady states, which
were reached after long spin-up periods; see the appendix for
details. The northern and southern hemispheres in the simulations are
statistically identical, so differences between the hemispheres in
figures showing long-term averages are indicative of the sampling
variability of the averages.

\section{Simulation results}
\label{s:Simulations}

\subsection{Upper-atmospheric zonal flow}

Figure~\ref{fig_flow_lat} shows the simulated mean zonal velocities
near the levels at which cloud features from which the observed flows
are inferred are suspected to occur: in the Jupiter simulation at
0.75~bar, corresponding to the layer of ammonia ice clouds on the
actual planet \citep{Atreya99}; in the Saturn simulation at 0.1~bar,
in a layer of tropospheric (e.g., ammonia) hazes
\citep{Sanchez-Lavega07}; in the Uranus and Neptune simulations at
25~mbar, near the top of the stratospheric layers in which
hydrocarbons would condense and form hazes \citep{Gibbard03}.

The simulations reproduce large-scale features of the observed flows
in the upper atmosphere (Fig.~\ref{fig_flow_lat}). The Jupiter and
Saturn simulations exhibit equatorial superrotation, the Uranus and
Neptune simulations equatorial subrotation. The equatorial jet in the
Jupiter simulation has similar strength ($\about
150\,\mathrm{m\,s^{-1}}$) and width as the observed jet. The
equatorial jet in the Saturn simulation is stronger ($\about
230\,\mathrm{m\,s^{-1}}$) and slightly wider than that in the Jupiter
simulation, but it is weaker and narrower than the observed jet at a
corresponding level on Saturn.  The Jupiter and Saturn simulations
exhibit alternating off-equatorial jets; they are broader than the
observed jets but of similar strength.  Especially in the Saturn
simulation, the retrograde jets (except for the first retrograde jet
off the equator) are broad and weak with speeds less than $10\,
\mathrm{m \, s^{-1}}$.  They are more manifest as local minima of the
zonal velocity than as actual retrograde jets. The Uranus and Neptune
simulations exhibit a single off-equatorial jet in each
hemisphere. The overall structure of the jets in the Uranus and
Neptune simulations is roughly consistent with observations, but the
equatorial jet in the Neptune simulation at the level shown ($\about
-40\,\mathrm{m\,s^{-1}}$) is considerably weaker than that
observed. (However, the jet is stronger at higher levels in the
simulation; see Fig.~\ref{fig_circulation} below.)

\begin{figure}[!htb]
  \centerline{\includegraphics{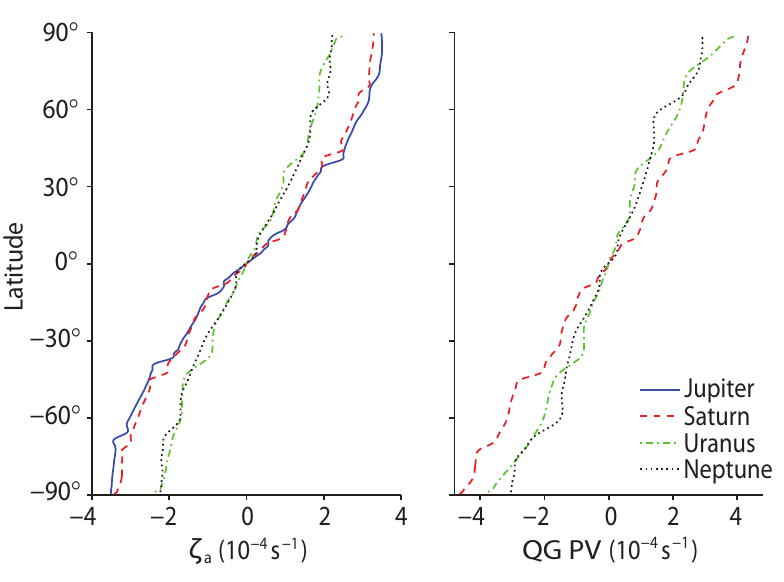}}
  \caption{Mean absolute vorticity (left) and quasigeostrophic
    potential vorticity (right) in the simulations, evaluated at the
    same levels at which the zonal velocities are shown in
    Fig.~\ref{fig_flow_lat}. The quasigeostrophic potential vorticity
    is calculated analogously to \citet{Read09b}. It is not shown for
    the Jupiter simulation because the stratification at the
    corresponding level is nearly statically neutral (cf.\
    Fig.~\ref{fig_circulation} below), so potential vorticity is not
    well defined.}
\label{f:abs_vort}
\end{figure}

In general, the prograde jets (or zonal velocity maxima) are sharper
than the retrograde jets (or zonal velocity minima), consistent with
the zonal velocity maxima being barotropically more stable
\citep{Rhines94}. Indeed, meridional gradients of both absolute
vorticity and quasigeostrophic potential vorticity at the levels at
which the zonal velocities are shown in Fig.~\ref{fig_flow_lat} are
small near zonal velocity minima and are reversed near some of them
(Fig.~\ref{f:abs_vort}). (The quasigeostrophic potential vorticity is
not shown for the Jupiter simulation because the stratification at the
corresponding level is nearly statically neutral, so that potential
vorticity is not well defined; see Fig.~\ref{fig_circulation} below.)
The changing magnitude of the vorticity gradients between zonal
velocity maxima and minima gives rise to a staircase pattern of
absolute vorticity and potential vorticity as a function of latitude
\citep{McIntyre82,Dritschel08}. Absolute vorticity gradients can reach
about $-2\beta$, particularly in higher latitudes in the flanks of the
minima; quasigeostrophic potential vorticity gradients are also
reversed near some of the zonal velocity minima, particularly in the
Uranus and Neptune simulations, but they do not reach as strongly
negative values as the absolute vorticity gradients. These features
are roughly consistent with observations of Jupiter and Saturn
\citep{Ingersoll81,Read06}, but quasigeostrophic potential vorticity
gradients may be more strongly negative on Saturn than they are in our
simulation \citep{Read09b}.  The vorticity profiles indicate that
barotropic instability limits the sharpening of the retrograde jets,
though not to the degree that the statistically steady states of the
flows would satisfy sufficient conditions for linear barotropic
stability for unforced and non-dissipative flows. (There is no reason
that such sufficient conditions ought to be satisfied in
forced-dissipative flows.)

\begin{figure}[!htb]
\centerline{
\includegraphics{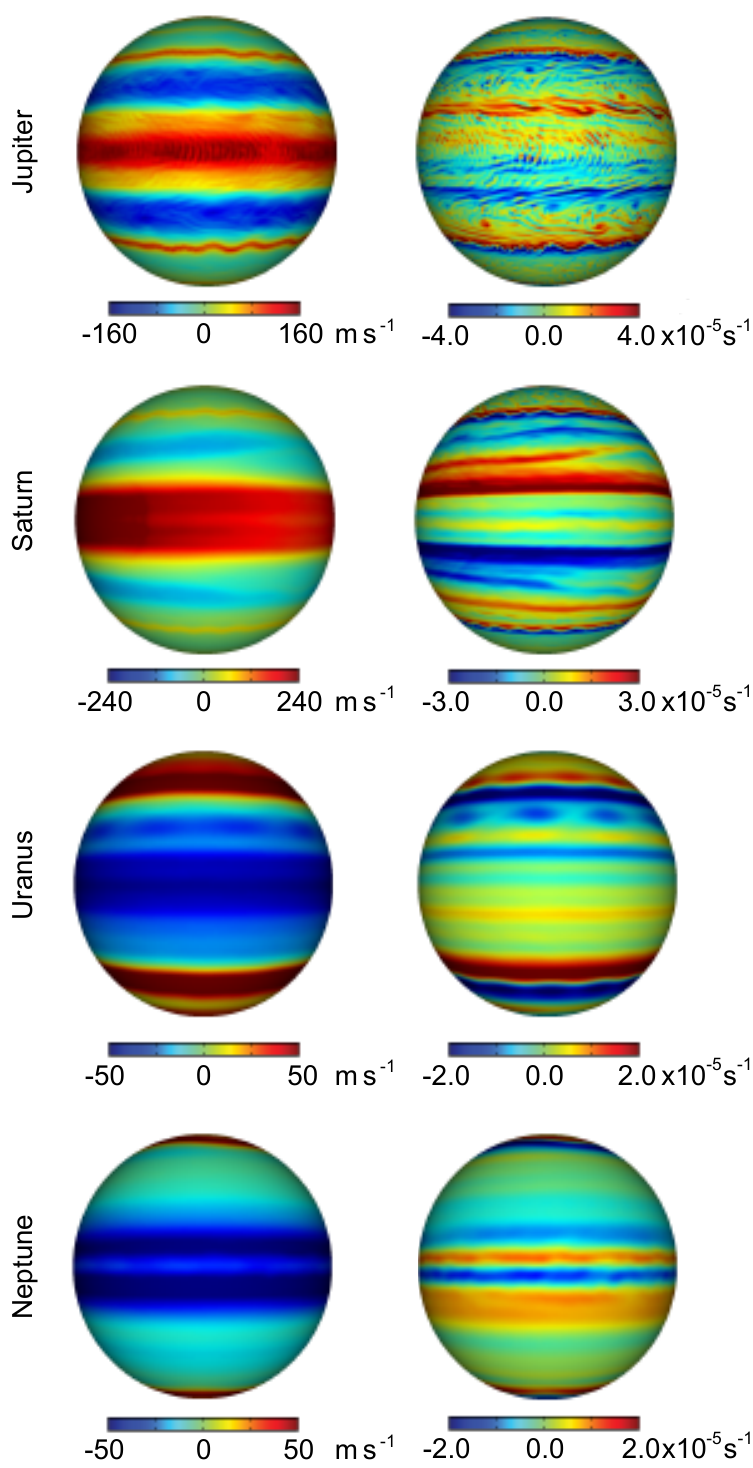}}
\caption{Zonal velocity (left column) and relative vorticity (right
  column) at one instant in the statistically steady state of the
  simulations. The levels at which the flow fields are shown are the
  same as in Fig.~\ref{fig_flow_lat}.  The equatorial Rossby waves
  (organized into large wave packets) that are responsible for the
  generation of the equatorial superrotation are recognizable in the
  Jupiter and Saturn simulations. Coherent vortices are clearly seen
  in the Jupiter and Uranus simulations.  Animations of the flow
  fields are available at
  http://www.gps.caltech.edu/\~{}tapio/pubs.html.}
\label{fig_vorticity_all}
\end{figure}

The jets are not only evident in long-term averages but also in
instantaneous flow fields. The instantaneous zonal velocity and
vorticity fields show the jets as well as large-scale jet undulations,
waves, and coherent vortices (Fig.~\ref{fig_vorticity_all}). In the
equatorial region in the Jupiter and Saturn simulations, the waves are
organized into large wave packets (Fig.~\ref{fig_vorticity_all}, left
column). Animations of the flow fields\footnote{available at
  http://www.gps.caltech.edu/\~{}tapio/pubs.html} show that the wave
packets exhibit westward group propagation,
as expected for long equatorial Rossby waves (\citealp{Matsuno66};
\citealp[chapter~11]{Gill82}). In the vorticity fields
(Fig.~\ref{fig_vorticity_all}, right column), coherent vortices are
seen in latitude regions with nearly homogenized absolute vorticity or
potential vorticity, that is, in regions with large negative curvature
of the zonal flow with latitude (cf.\ Fig.~\ref{f:abs_vort}).

\subsection{High-latitude coherent vortices and waves on jets}

\begin{figure}[!htb]
  \centerline{\includegraphics{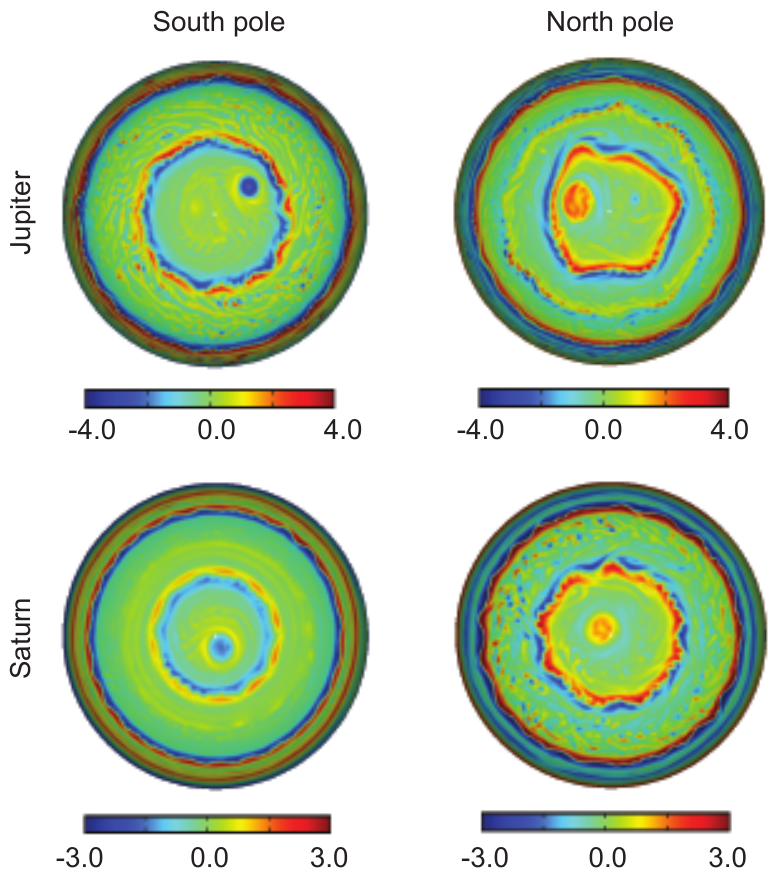}}
  \caption{Relative vorticity ($10^{-5} \,\mathrm{s^{-1}}$) in high
    latitudes at one instant in the statistically steady state of the
    Jupiter and Saturn simulations. Left column: south polar
    projection; right column: north polar projection. The vorticity is
    shown at the same levels as the flow fields in
    Figs.~\ref{fig_flow_lat} and \ref{fig_vorticity_all} (0.75 and
    0.1~bar, respectively).}
\label{fig_large_vortex}
\end{figure}

In high latitudes in the Jupiter and Saturn simulations, very large
coherent vortices ($20^{\circ} \times 10^{\circ}$
longitude$\times$latitude) form spontaneously
(Fig.~\ref{fig_large_vortex}). They extend all the way to the bottom
of the domain, with the magnitude of the vorticity decreasing weakly
with depth: the peak vorticity at the bottom of the domain is about
$80 \%$ of its maximum value in the column.

The large coherent vortices are cyclonic, with typical vorticities of
magnitude $\about 2 \times 10^{-5} \ \mathrm{s^{-1}}$. They are
advected by the flow in their environment and have a local temperature
minimum at the center ($\about 10\,\mathrm{K}$ lower temperature than
the environment). These cyclonic vortices are long-lived, with
life-spans apparently determined by the radiative timescale (the
timescale on which eddies can modify the mean flow). In the Jupiter
simulation with an atmosphere of $3$~bar thickness, the radiative
timescale is $\about 10$ Earth years; it is $\about 50$ Earth years in
the Saturn simulation. Since the radiative timescale increases with
pressure, it is longer for deeper atmospheres, which might explain why
the observed coherent vortices such as the Great Red Spot on Jupiter
are so long-lived.

Coherent vortices preferentially exist in regions where absolute
vorticity or potential vorticity gradients vanish, as they can then
arise spontaneously in barotropic or quasigeostrophic flows and remain
stable \citep[e.g.,][]{McWilliams84, marcus88, marcus93}. Since the
planetary vorticity gradient vanishes at the poles, formation of
coherent vortices in high latitudes may require less vorticity mixing
in the environment than it does at lower latitudes. Hence, the large
coherent vortices in high latitudes may appear earlier in
simulations. If the simulations were conducted for a (much) longer
period and if numerical (subgrid-scale) dissipation could be further
reduced, it is possible that large coherent vortices would also appear
in lower latitudes, such as the latitude ($23^\circ$S planetocentric)
of the Great Red Spot on Jupiter, which is embedded in an environment
of small absolute vorticity and potential vorticity gradients
\citep{Ingersoll81,Read06}.

The coherent polar vortices in our simulations are contained in the
polar cap bounded by the highest-latitude prograde jet. These polar
jets exhibit large-scale undulations, as do other jets (cf.\
Fig.~\ref{fig_vorticity_all}). In the Saturn simulation, for example,
the prograde polar jet at $68^\circ$N exhibits a wavenumber-8 or 9
undulation (Fig.~\ref{fig_large_vortex}), with a zonal phase velocity
of $-24\,\mathrm{m\,s^{-1}}$. This is retrograde, and retrograde
relative to the mean flow,
consistent with the undulation being a Rossby wave. The undulation is
reminiscent of the nearly stationary wavenumber-6 pattern (``polar
hexagon'') observed in Saturn's polar atmosphere at $76^\circ$N
planetocentric latitude \citep{Godfrey88,Allison90,Fletcher08}.
Indeed, with a smaller drag coefficient that leads to a slightly
stronger polar jet (see section~\ref{s:Mechanisms}\ref{s:drag}), we
also obtain a wavenumber-6 pattern on the polar jet in our
simulations; we will describe this in greater detail elsewhere.

\subsection{Vertical structure of zonal flow}

\begin{figure*}[!htb]
\centering\includegraphics{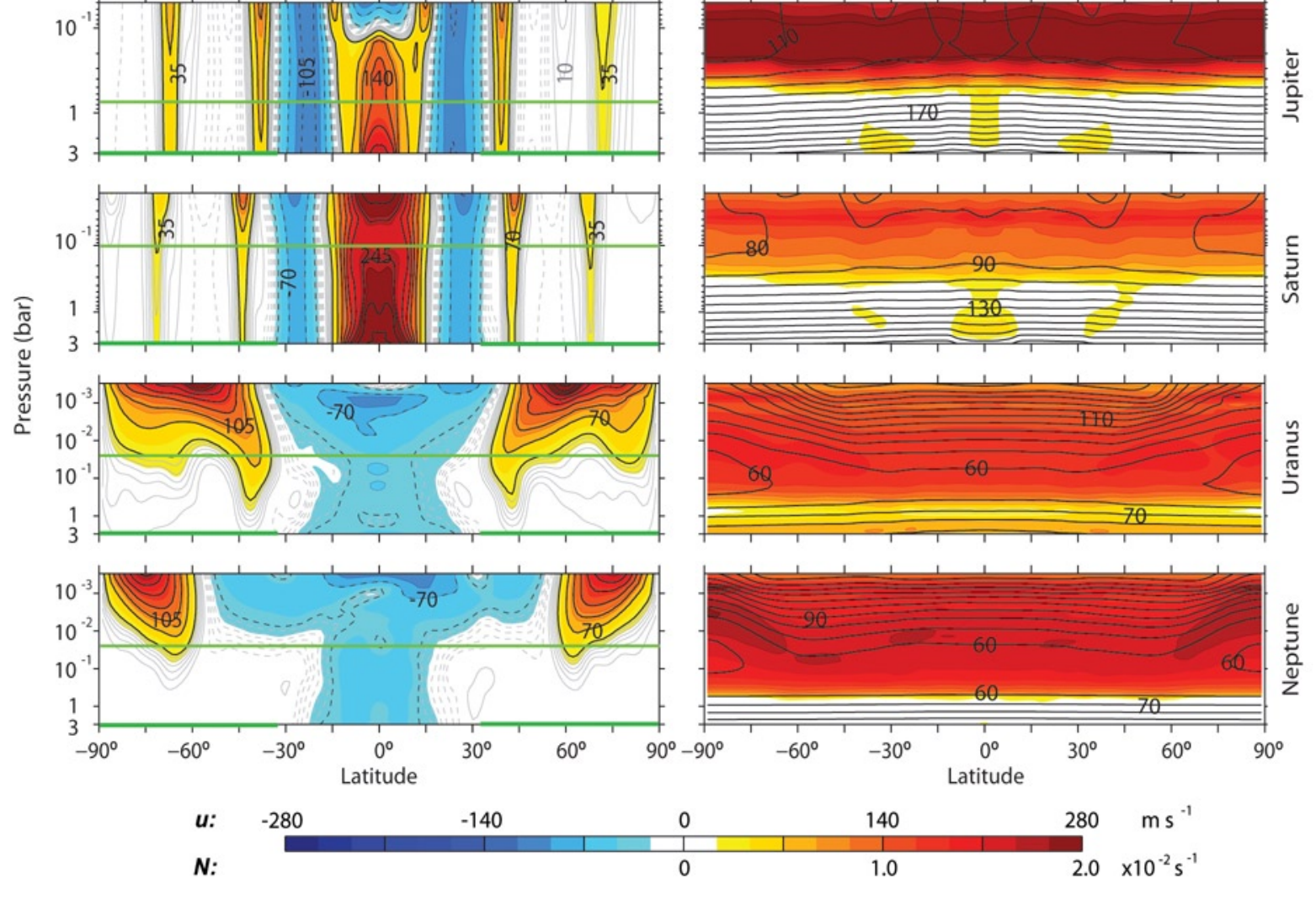}
\caption{Mean flow fields in the latitude-pressure plane in the
  simulations. The left column shows zonal-flow profiles: Gray
  contours for zonal flow speeds between $5$ and $30 \, \mathrm{m \,
    s^{-1}}$, with a contour interval of $5 \, \mathrm{m \, s^{-1}}$;
  black contours for zonal flow speeds of $35 \, \mathrm{m \, s^{-1}}$
  or above, with a contour interval of of $35 \, \mathrm{m \,
    s^{-1}}$. Solid contours and red tones for prograde flow and
  dashed contours and blue tones for retrograde flow. The right column
  shows temperature (contours, contour interval 10~K) and buoyancy
  frequency $N$ (colors). The thick green parts of the latitude axes
  in the left column mark the latitudes with nonzero drag. The thin
  green lines indicate the levels at which flow fields are shown in
  Figs.~\ref{fig_flow_lat}--\ref{fig_large_vortex}.}
\label{fig_circulation}
\end{figure*}

The simulated flows in
Figs.~\ref{fig_flow_lat}--\ref{fig_large_vortex} were shown near the
suspected levels of observed cloud features on the giant
planets. However, the flows in the simulations vary in the
vertical. The prograde equatorial jets in the Jupiter and Saturn
simulations strengthen with depth (Fig.~\ref{fig_circulation}, left
column).  The corresponding vertical shear of the zonal flow ($\about
1$--$2 \times 10^{-3} \, \mathrm{s^{-1}}$) is similar to that measured
by the \emph{Galileo} probe on Jupiter between 0.7~bar and 4~bar
\citep{Atkinson98} and to that inferred from \emph{Cassini} data for
Saturn between 0.05 and 0.8~bar (\citealp{Sanchez-Lavega07}; see also
the zonal flow observations at different levels in
Fig.~\ref{fig_flow_lat}).  The retrograde equatorial jets in the
Uranus and Neptune simulations are strongest in the stratosphere and
weaken with depth, consistent with inferences drawn from gravity
measurements with the \emph{Voyager~2} spacecraft
\citep{Hubbard91}. Away from the equator, prograde jets generally
weaken with depth and retrograde jets strengthen slightly or do not
vary much with depth.

\subsection{Temperature structure}

Consistent with thermal wind balance, temperatures increase
equatorward along isobars where prograde jets weaken with depth or
retrograde jets strengthen with depth, and they decrease equatorward
where the opposite is true. Therefore, in the equatorial upper
troposphere in the Jupiter and Saturn simulations, where the prograde
jets strengthen with depth, temperatures decrease equatorward and have
a minimum at the equator (Fig.~\ref{fig_circulation}, contours in
right column). A similar equatorial temperature minimum is seen in
observations of Jupiter and Saturn
\citep{Simon-Miller06,Fletcher07}. The tropopause, recognizable as the
level at which the vertical temperature lapse rate changes sign, in
all simulations lies near $0.1$~bar, likewise as observed
\citep{Simon-Miller06, Fletcher07}.  Below the tropopause,
temperatures increase with depth. In the Jupiter, Saturn, and Neptune
simulations, the atmosphere is close to statically neutrally
stratified below the statically stable layer near the tropopause
because of vigorous convection driven by intrinsic heat fluxes. In the
Uranus simulation, the entire atmosphere is stably stratified (or
close to it) because convection and intrinsic heat fluxes are weak
(Fig.~\ref{fig_circulation}, colors in right column).

In the Jupiter simulation, the equator-to-pole contrast in the
brightness temperature of thermal radiation is 10~K, which is similar
to albeit larger than the observed brightness temperature contrast or
the observed temperature contrast near the emission level
\citep{Ingersoll76, Ingersoll90, Ingersoll04,
  Simon-Miller06}.\footnote{Variations in brightness temperature
  gradients off the equator may be weaker in the Jupiter and Saturn
  simulations than they are in observations, at least at some
  wavelengths \citep{Ingersoll90,Ingersoll04}.}  In the Saturn
simulation, the equator-to-pole brightness temperature contrast is
7~K, consistent with observations \citep{Ingersoll90,Fletcher07}. In
the Uranus and Neptune simulations, the equator-to-pole brightness
temperature contrasts are 8~K and 2~K, respectively, consistent with
observations \citep{Ingersoll90}. That is, meridional enthalpy
transport in all simulations substantially reduces the (much greater)
radiative-convective equilibrium temperature contrasts near the
emission levels. (The emission levels lie between about $0.3$ and
$0.5$~bar in our simulations, and radiative-convective equilibrium
temperature contrasts there vary between 20~K for Neptune and 35 K for
Uranus.) It does not appear necessary to invoke meridional mixing deep
in the atmosphere to account for the smallness of the observed
brightness temperature contrasts
\citep[cf.][]{Ingersoll76b,Ingersoll78}.

\begin{figure}[!htb]
  \centering%
  \includegraphics{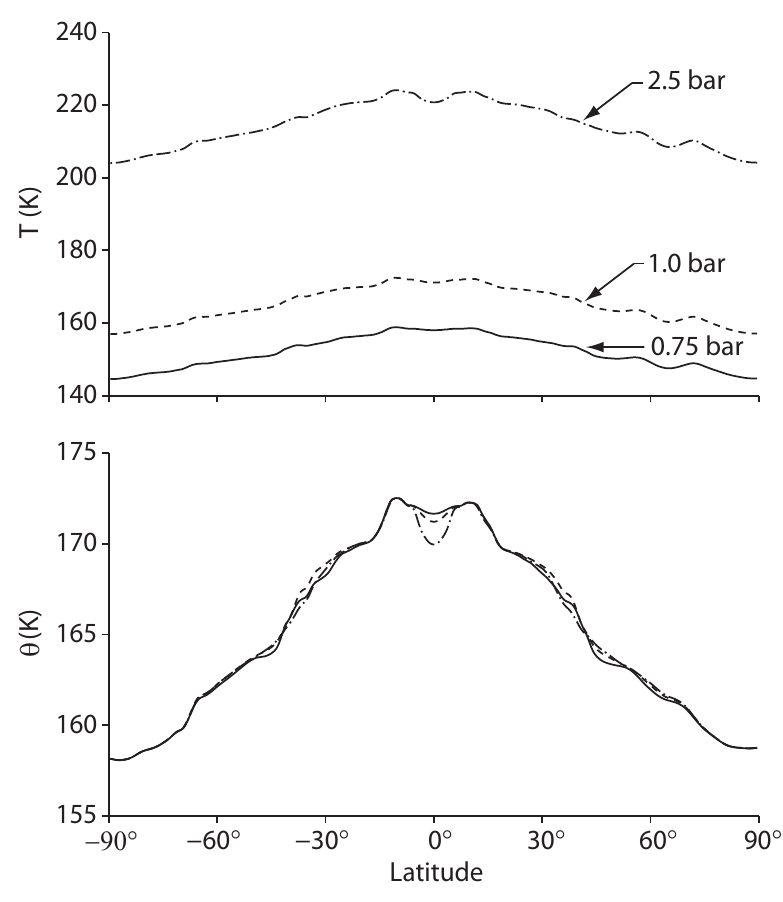}
  \caption{Temperature (a) and potential temperature (b) at the
    $0.75$-bar (solid), $1.0$-bar (dashed), and $2.5$-bar
    (dash-dotted) levels in Jupiter simulation. The potential
    temperature $\theta = T (p_0/p)^{\kappa}$ is evaluated with the
    reference pressure $p_0=1$~bar.}
\label{fig:pottemp_temp}
\end{figure}

However, although temperature contrasts at the emission level are
generally small, equator-to-pole temperature contrasts at higher or
lower levels of the simulated atmospheres differ, and the same is
likely true for the actual planets. For example, while temperature
contrasts near the emission level in the Jupiter simulation are small,
they are greater at lower levels where temperatures are greater
(Fig.~\ref{fig:pottemp_temp}a). Because the atmosphere is close to
statically neutrally stratified below the upper troposphere, entropy
(potential temperature) there is constant in the vertical. [More
generally, entropy is constant along angular momentum surfaces, to
achieve a state of neutrality with respect to slantwise convection;
see \citet{Emanuel83b} and \citet{Thorpe89}.] Hence, the meridional
potential temperature distribution at lower levels is the same as that
near the top of the neutrally stratified layer, except near the
equator where the atmosphere has a weak positive static stability
(Fig.~\ref{fig:pottemp_temp}b). But this implies that meridional
temperature gradients off the equator increase with pressure, as
temperature $T$ and potential temperature $\theta$ are related by $T =
\theta (p/p_0)^\kappa$, where $p$ is pressure and $p_0$ a constant
reference pressure. In particular, the signatures of the prograde
off-equatorial jets weakening with depth (enhanced meridional
temperature gradients) and of the retrograde off-equatorial jets
strengthening slightly or not varying with depth (reduced or vanishing
meridional temperature gradients) are visible at all levels
(Figs.~\ref{fig:pottemp_temp}a and b). There are no observations of
entropy or temperature distributions below the upper troposphere for
the giant planets, but we expect them to behave similarly as in our
simulations, for the reasons discussed in
section~\ref{s:AM_balance}\ref{s:temperature} below.

\section{Mechanisms of jet formation}
\label{s:Mechanisms}

Why are the flow and temperature structures in the giant planet
simulations so different? The fundamental reason lies in the different
strengths of the differential radiative heating and the intrinsic heat
flux, and in the different ways in which the two can lead to the
generation of the eddies that maintain the jets. Eddies in rapidly
rotating atmospheres generally transport angular momentum from their
dissipation (breaking) region into their generation region
(\citealp{Held75,Andrews76,Andrews78c,Rhines94,Held00b};
\citealp[][chapter~12]{Vallis06}). If they are preferentially
generated in prograde jets, they lead to angular momentum transport
from retrograde into prograde jets, which can maintain the jets
against dissipation \citep[e.g.,][]{Vallis06,OGorman08a}. Such angular
momentum transport from retrograde into prograde jets has indeed been
observed on Jupiter and Saturn
\citep{Ingersoll81,Salyk06,DelGenio07}. A central question is, then,
what kind of eddies can give rise to the angular momentum transport
required to spin up and maintain the jets? We have addressed this
question more formally and in greater detail in SL09. Here we
summarize some results from that earlier paper and expand on some that
are important for understanding the simulations presented in this
paper.

\subsection{Off-equatorial jets}

\begin{figure*}[!tbh]
\centering\includegraphics{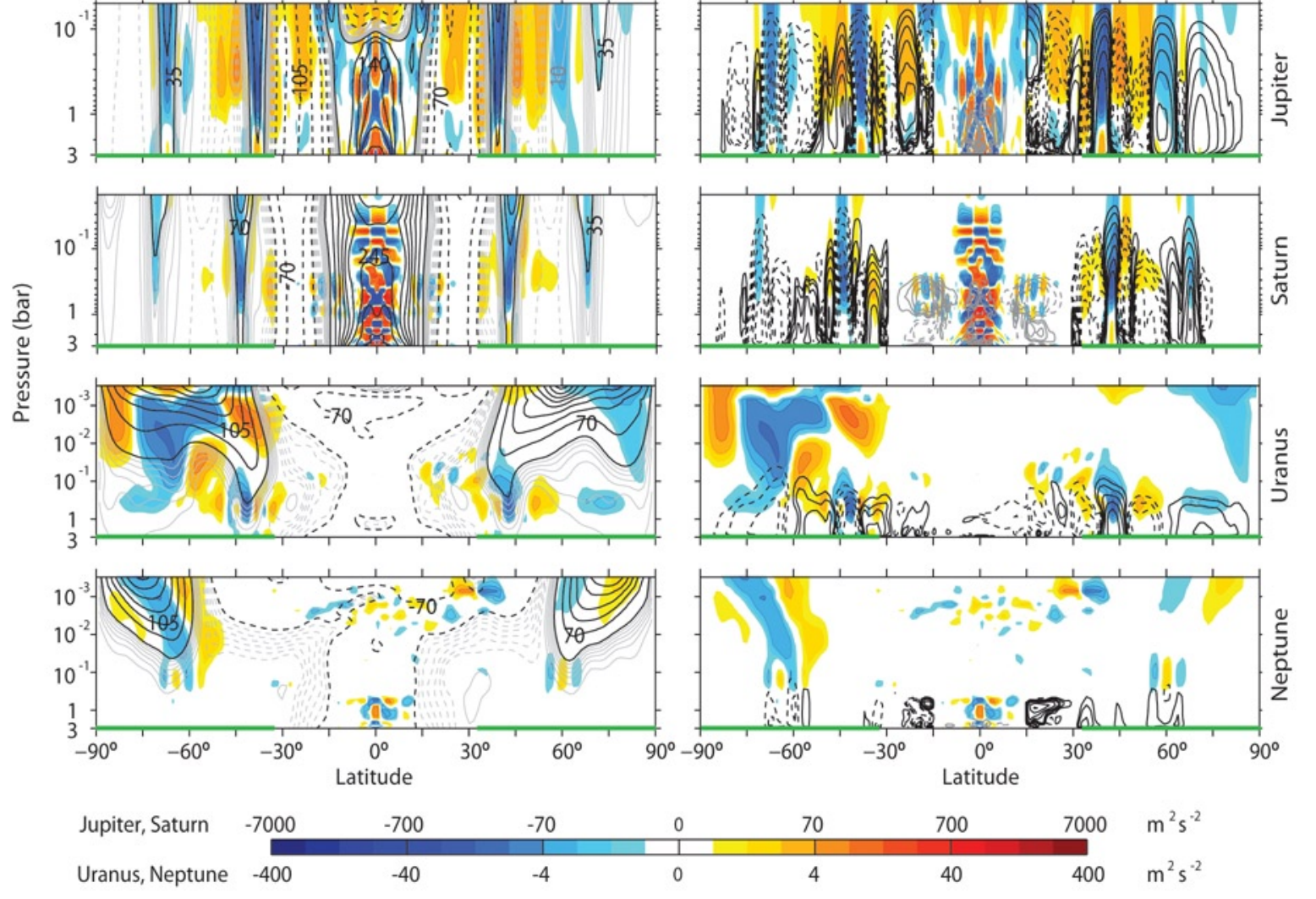}
\caption{Mean zonal velocities, mass flux streamfunction (contours)
  and divergence $\divy (\overline{u'v'} \, a \cos\phi)$ of meridional
  eddy angular momentum fluxes (colors) in the simulations. The left
  column shows zonal-flow profiles (contours as in
  Fig.~\ref{fig_circulation}) and eddy angular momentum flux
  divergence (colors). The right column shows the mass flux
  streamfunction (contours) and the same eddy angular momentum flux
  divergence as in the left column (colors). The contouring for the
  streamfunction is logarithmic: black contours from $\pm 1$ to $\pm
  64 \times 10^8 \,\mathrm{kg\,s^{-1}}$; gray contours for absolute
  values greater than or equal to $128 \times
  10^{8}\,\mathrm{kg\,s^{-1}}$, with factors of 2 separating contour
  levels. (Black contours that would be within gray contours are not
  shown.) Solid contours for positive streamfunction values
  (counterclockwise rotation) and dashed contours for negative
  streamfunction values (clockwise rotation). Some streamfunction
  contours are truncated at the bottom of the plotting domain (3~bar)
  because they close at higher pressures (the pressure at the bottom
  of the GCM domain can locally exceed 3~bar). The contouring for the
  eddy angular momentum flux divergence is likewise logarithmic, with
  the scale shown in the colorbar. As in Fig.~\ref{fig_circulation},
  the thick green parts of the latitude axes mark the latitudes with
  nonzero drag.}
\label{fig_streamfunction}
\end{figure*}

Away from the equator, the differential radiative heating of the upper
atmospheres produces meridional temperature gradients, which are
baroclinically unstable and lead to eddy generation. Eddy generation
preferentially occurs in the troposphere in the baroclinically more
unstable prograde jets with enhanced temperature gradients and
enhanced prograde vertical shear (Fig.~\ref{fig_circulation}). It
results in angular momentum transport from retrograde into prograde
tropospheric jets (Fig.~\ref{fig_streamfunction}). This angular
momentum transport maintains the off-equatorial tropospheric jets in
the Jupiter and Saturn simulations against dissipation at depth. It
has a baroclinic structure and, in the Jupiter and Saturn simulations,
is in structure and magnitude consistent with observations
\citep[cf.][]{Ingersoll81,Salyk06,DelGenio07}. The conversion rate of
eddy to mean-flow kinetic energy is of order $10^{-5}
\,\mathrm{W\,m^{-3}}$ in the upper tropospheres in the Jupiter and
Saturn simulations---as observed. In the global mean, the conversion
rates are $0.09 \,\mathrm{W\,m^{-2}}$ and $0.026 \,\mathrm{W\,m^{-2}}$
in the Jupiter and Saturn simulations, respectively, implying that the
conversion rate in either simulation is about 0.6\% of the total
energy uptake of the atmosphere. Consistent with a baroclinic eddy
generation mechanism, off-equatorial eddy angular momentum fluxes and
jets disappear in a Jupiter control simulation in which baroclinic
instability is suppressed by imposing insolation uniformly at the top
of the atmosphere (SL09).

The off-equatorial jets in the Uranus and Neptune simulations are
situated in the stratosphere and are broader than those in the Jupiter
and Saturn simulations. They appear to be broader because only the
longest waves generated at lower levels are able to reach the
stratosphere \citep{Charney61}. Indeed, the eddy angular momentum flux
divergence at the stratospheric jet cores is much weaker than that at
the tropospheric jet cores in the Jupiter and Saturn simulations
(Fig.~\ref{fig_streamfunction}). It is not even of a consistent sign
at all times but exhibits considerable low-frequency variability, as
indicated by the differences between the statistically identical
hemispheres in the long-term (1500-day) averages shown in
Fig.~\ref{fig_streamfunction}. The jets become very weak below the
tropopause and, particularly in the Uranus simulation, give way to a
tropospheric zonal flow with smaller meridional scales
(Fig.~\ref{fig_circulation}). The structure of the stratospheric jets
implies that they interact only weakly with the drag at the lower
boundary, so only weak eddy angular momentum flux divergence is
necessary to maintain them. The jets are primarily a manifestation of
the thermal structure and of the vertical shear of the zonal flow
implied by it. The latter do not exhibit smaller-scale variations
because smaller-scale eddy transports of angular momentum and heat in
the stratosphere are weak.

\subsection{Equatorial superrotation}

Near the equator, convection can penetrate into the upper troposphere
and can generate Rossby waves if the intrinsic heat flux is strong
enough to overcome the static stabilization of the atmosphere by the
radiative heating from above. Fluctuations in convective heating are
primarily balanced by vertical motion and, at the level of the
convective outflows, by horizontal divergence of mass fluxes, as in
the tropics of Earth's atmosphere. That is, the dominant balance in
the thermodynamic equation is the weak-temperature gradient (WTG)
balance \citep{Sobel01},\footnote{The WTG approximation holds where
  the Rossby number satisfies $\mathrm{Ro}=U/|fL| \gtrsim 1$ and the
  Froude number satisfies $\mathrm{Fr}=UV/(gH) \ll 1$, where $U$ is a
  zonal velocity scale, $V$ a meridional or eddy velocity scale, $L$ a
  length scale of flow variations, and $H$ the scale height
  (\citealp{Charney63}; SL09). In SL09, we showed that the WTG
  approximation holds within $\about 4^\circ$ of the equator in
  Jupiter's upper troposphere. Analogous scale analysis suggests the
  WTG approximation holds within $\about 7^\circ$ of the equator in
  Saturn's upper troposphere. For Uranus' and Neptune's upper
  tropospheres, no flow data are available, but with the tropospheric
  velocity scales from our simulations, the WTG approximation holds
  within $\about 5^\circ$ of the equator.\label{fn:Ro}}
\begin{equation}\label{e:wtg_thermo}
  \divh \vec{v}_\chi \approx \partial_p (Q/S).
\end{equation}
Here, $\vec{v}_\chi$ denotes the divergent horizontal flow component,
$Q=D\theta/Dt$ the diabatic heating rate, and $S=-\partial_p \theta$
the static stability; the subscript $h$ on the differential operator
$\vec{\nabla}_h$ signifies horizontal derivative. As discussed in
\citet{Sardeshmukh88}, fluctuations in the horizontal divergence are a
source of equatorial Rossby waves, and the fluctuating vorticity
source
\begin{subequations}\label{e:rossby_source}
\begin{equation}
  R' = R - \Bar R,
\end{equation}
with
\begin{equation}
  R = -\zeta_a (\divh \vec{v}_\chi) - (\vec{v}_\chi \cdot \gradh) \zeta_a,
\end{equation}
\end{subequations}
can be taken to be the Rossby wave source. (The overbar denotes the
isobaric zonal and temporal mean and primes deviations therefrom.)
Thus, fluctuations in convective heating lead to horizontal divergence
fluctuations \eqref{e:wtg_thermo}, which can generate equatorial
Rossby waves through stretching of absolute vorticity $\zeta_a$ or
advection of absolute vorticity by the divergent flow $\vec{v}_\chi$
\eqref{e:rossby_source}. Because the planetary vorticity vanishes at
the equator, the Rossby wave source $R'$ typically has largest
amplitude just off the equator, but it does not necessarily vanish at
the equator or where the absolute vorticity vanishes because absolute
vorticity advection by the divergent flow may not vanish.

Equatorial Rossby waves, organized into large-scale wave packets, are
recognizable in the Jupiter and Saturn simulations in
Fig.~\ref{fig_vorticity_all}. In the Jupiter simulation, the
energy-containing zonal wavenumber is $\about 10$, corresponding to
the wavenumber of the wave packet envelope. Waves with similar scales
have also been observed on Jupiter \citep{Allison90}.  The retrograde
tilt of the waves' phase lines away from the equator, clearly seen in
the Jupiter simulation (Fig.~\ref{fig_vorticity_all}), indicates that
they transport angular momentum toward the equator
\citep[cf.][chapter~11]{Peixoto92}. This is generally to be expected
for such convectively generated Rossby waves: they transport angular
momentum toward the equatorial region because this is where they are
preferentially generated. The Rossby wave source $R'$ owing to
horizontally divergent flow has largest amplitude in the equatorial
region because only there will convective heating fluctuations
necessarily lead to horizontal divergence fluctuations on large
scales; away from the equator, the WTG approximation
\eqref{e:wtg_thermo} of the thermodynamic equation does not hold. [See
SL09 (their Fig.~5) for a demonstration that $R'$ has largest
amplitude near the equator.]  The angular momentum transport toward
the equatorial region by convectively generated Rossby waves leads to
equatorial superrotation if it is sufficiently strong and drag on the
zonal flow is sufficiently weak (SL09).

Convective Rossby wave generation near the equator is the key process
responsible for superrotation in the Jupiter and Saturn
simulations. In SL09, we demonstrated that without intrinsic heat
fluxes and the convection they induce, a Jupiter simulation similar to
the one here exhibits equatorial subrotation; the same is true for the
Jupiter and Saturn simulations here. Therefore, we suggest convective
Rossby wave generation is what causes the superrotating equatorial
jets on Jupiter and Saturn.

When convective Rossby wave generation produces equatorial
superrotation, it produces a jet whose half-width $L_s$ is similar to
the scale of equatorial Rossby waves: the equatorial Rossby radius
$L_\beta = \sqrt{c/\beta}$, with gravity wave speed $c$ and planetary
vorticity gradient $\beta$.  Vorticity mixing arguments give an
estimate for the maximum strength of the equatorial jet
(\citealp{Rhines94}, SL09): If the end state of vorticity mixing is a
state in which the absolute vorticity is homogenized across the
equatorial jet in each hemisphere separately, with a barotropically
stable jump at the equator, and if the jet half-width $L_s$ is similar
to the equatorial Rossby radius $L_\beta$, the jet speed at the
equator will be
\begin{equation}
  U \lesssim \frac{\beta L_s^2}{2} \sim \frac{c}{2}.
\end{equation}
To the extent that this bound is attained (at the level of maximum
equatorial jet speed), the jet speed increases quadratically with the
jet width. This is roughly consistent with observations of Jupiter and
Saturn (though the maximum equatorial jet speed on Saturn is not known
for lack of observations deeper in the atmosphere). It is also roughly
consistent with our simulations, although a state of homogenized
absolute vorticity in the equatorial region of each hemisphere is not
attained in the simulations. That is, the equatorial jet on Saturn may
be stronger and wider than that on Jupiter because the gravity wave
speed is larger.

The flow configurations in the Jupiter and Saturn simulations differ
qualitatively from those in the Uranus and Neptune simulations because
the relative strengths of baroclinic eddy generation away from the
equator and convective Rossby wave generation near the equator differ.
In the Uranus simulation, the intrinsic heat flux is negligible, the
atmosphere is stably stratified, and there is no substantial
convective Rossby wave source near the equator. Consequently, the
equatorial eddy kinetic energy is weak, and the equatorial flow is
retrograde.

In the Neptune simulation, the intrinsic heat flux is strong enough
that convection penetrates into the upper troposphere. As in the other
simulations, eddies can be generated by (a) baroclinic instability off
the equator induced by differential solar heating, or (b) convective
Rossby wave generation near the equator induced by the intrinsic heat
flux. Eddies produced by these two mechanisms compete with each other
in their contribution to the angular momentum transport to or from low
latitudes.  Off-equatorial baroclinic eddy generation implies angular
momentum flux convergence in the off-equatorial generation regions and
divergence in lower latitudes, and hence a tendency toward retrograde
equatorial flow. Convective Rossby wave generation near the equator
can lead to prograde equatorial flow, but in the Neptune simulation,
the rms Rossby wave source $R'$ in the equatorial region is much
smaller than that in the Jupiter and Saturn simulations: the rms
Rossby wave source $R'$ in the upper troposphere near the equator is
$\about 10^{-12} \, \mathrm{s^{-2}}$ for Neptune but $\about 10^{-10}
\, \mathrm{s^{-2}}$ for Jupiter and Saturn. Convective Rossby wave
generation and the associated angular momentum flux convergence near
the equator appear to be too weak to overcome the angular momentum
flux divergence in low latitudes that is caused by eddies generated
baroclinically away from the equator. As a consequence, the equatorial
flow is retrograde. This is corroborated by control simulations.

\subsection{Neptune control simulations}

We investigated the relative roles of baroclinic eddy generation and
convective Rossby wave generation on Neptune in two control
simulations, one in which the convective Rossby wave source was
enhanced and one in which the baroclinic eddy generation caused by
differential solar heating was suppressed. Because angular momentum
flux divergence in low latitudes owing to baroclinic eddy generation
away from the equator can counteract any angular momentum flux
convergence owing to convective Rossby wave generation, generation of
equatorial superrotation in the Neptune simulation may require a
stronger intrinsic heat flux or weaker differential solar heating.

\begin{figure*}[!htb]
\centering\includegraphics{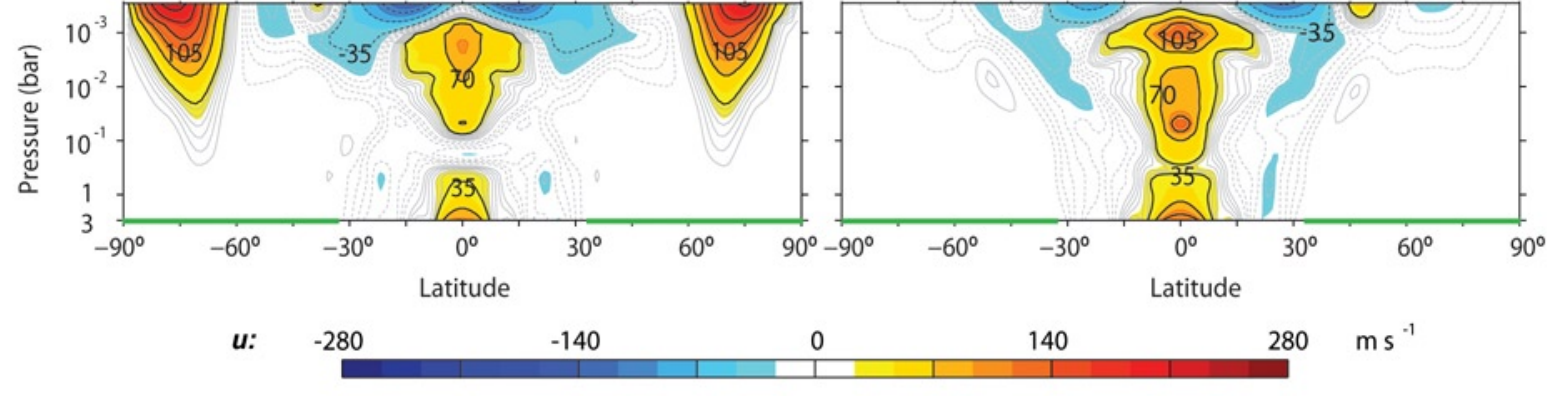}
\caption{Mean zonal velocities in the latitude-pressure plane in
  Neptune control simulations. Contour intervals and colors as in
  Fig.~\ref{fig_circulation}.  Left: simulation with Neptune's
  physical parameters but Saturn's intrinsic heat flux ($2.01 \
  \mathrm{W \ m^{-2}}$).  Right: simulation with Neptune's physical
  parameters but uniform insolation at the top of the atmosphere.}
\label{neptune}
\end{figure*}

Indeed, a control simulation with Neptune's physical parameters but in
which the convective Rossby wave source was enhanced by enhancing the
intrinsic heat flux---setting it to Saturn's $2.01 \, \mathrm{W \
  m^{-2}}$ in place of Neptune's $0.433\, \mathrm{W \
  m^{-2}}$---exhibits equatorial superrotation (Fig.~\ref{neptune},
left column). Conversely, a control simulation in which baroclinic
eddy generation was suppressed by imposing insolation uniformly at the
top of the atmosphere (but keeping the global mean fixed) also
exhibits equatorial superrotation (Fig.~\ref{neptune}, right
column). The prograde off-equatorial jets disappear with the
suppression of baroclinicity.

We suggest, then, that Uranus and Neptune exhibit equatorial
subrotation because baroclinic eddy generation away from the equator
is strong compared with convective Rossby wave generation near the
equator. Interestingly, our simulations suggest that Neptune's
intrinsic heat flux only needs to be larger by an $O(1)$ factor for
Neptune's atmosphere to develop equatorial superrotation, implying
that Neptune may have been superrotating earlier in its history, when
intrinsic heat fluxes were stronger.

\subsection{Effect of drag formulation on simulated
  flows}\label{s:drag}
One relatively unconstrained aspect of our simulations is the strength
and functional form of the drag at the artificial lower boundary. We
investigated the sensitivity of our results to the drag formulation by
varying it in a few Jupiter simulations.
\begin{figure*}[!htb]
\centerline{\includegraphics{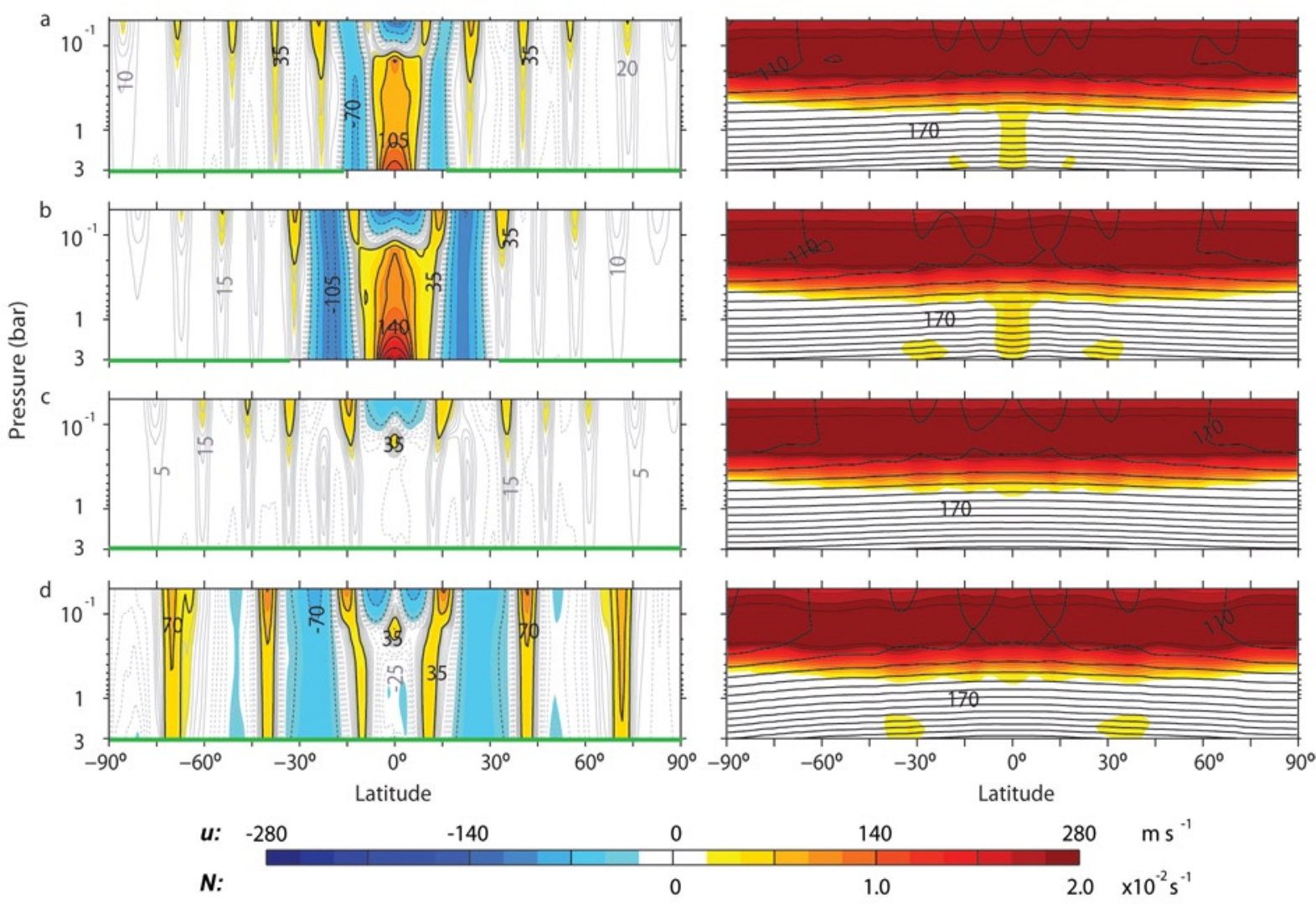}}
\caption{Mean flow fields in the latitude-pressure plane in Jupiter
  simulations with different drag formulations. The left column shows
  the zonal flow and the right column the temperature and buoyancy
  frequency, with the same plotting conventions and contour intervals
  as in Fig.~\ref{fig_circulation}.  (a) No drag in the equatorial
  region up to $\phi_0=16^\circ$, and off-equatorial drag coefficient
  $k_0 = 1/(20\,\mathrm{days})$. (b) No drag in the equatorial region
  up to $\phi_0=33^\circ$, and off-equatorial drag coefficient $k_0 =
  1/(10\,\mathrm{days})$. (c) Constant drag at all latitudes with drag
  coefficient $k_0 = 1/(10\,\mathrm{days})$. (d) Constant drag at all
  latitudes with drag coefficient $k_0 = 1/(100\,\mathrm{days})$. }
\label{fig_drag}
\end{figure*}

The Jupiter simulation in SL09 had an equatorial no-drag region half
as wide as that here (extending to $\phi_0=16^{\circ}$ latitude vs.\
$\phi_0=33^{\circ}$ here), in addition to having a larger drag
coefficient off the equator [$k_0 = 1/(20\,\mathrm{days})$ vs.\ $k_0 =
1/(100\,\mathrm{days})$ here]. Mean flow fields from that earlier
simulation are shown in Fig.~\ref{fig_drag}a. The half-width of the
prograde equatorial jet is $\about 5^{\circ}$ smaller than in the
Jupiter simulation reported here (cf.\
Fig.~\ref{fig_circulation}). The adjacent strong retrograde jets
appear to be confined to the no-drag region and hence do not extend as
far poleward as in the simulation with the wider no-drag region. The
off-equatorial jets are somewhat weaker and narrower---a result of the
enhanced off-equatorial drag, consistent with theories and other
simulations of geophysical turbulence
\citep{Smith02b,Danilov02}. However, the width of the equatorial
no-drag region does not primarily control the strength or width of the
equatorial jet, as evidenced by the relatively moderate changes in the
flow in low latitudes in response to the factor 2 change in the width
of the no-drag region.

That the strength and width of off-equatorial jets depend on the drag
coefficient is directly illustrated by simulations in which we
increased the off-equatorial drag coefficient further
[Fig.~\ref{fig_drag}b, $k_0 = 1/(10\,\mathrm{days})$]. The
off-equatorial jets become weaker and narrower as the drag coefficient
is increased. However, if the same enhanced drag is used at all
latitudes, without an equatorial no-drag region, there is no
large-scale prograde jet at the equator, while the off-equatorial flow
is not substantially modified (Fig.~\ref{fig_drag}c). (But a narrow
and shallow prograde jet forms near the tropopause at the equator.)
Similarly, if a weaker constant drag is used at all latitudes [$k_0 =
1/(100\,\mathrm{days})$], without an equatorial no-drag region, there
is likewise no large-scale prograde jet at the equator
(Fig.~\ref{fig_drag}d). Even weaker equatorial drag is required to
obtain a large-scale prograde jet at the equator if intrinsic heat
fluxes are specified consistent with observations. Consistent with the
theoretical arguments in SL09, stronger equatorial drag requires
larger intrinsic heat fluxes and thus a stronger equatorial Rossby
wave source to lead to superrotation. However, neither the precise
functional form of the drag, nor the magnitude of the drag coefficient
where it is nonzero, nor the width of the no-drag region appear to be
essential for our results---as long as there is an equatorial region
with no or sufficiently low drag such that a large-scale prograde jet
can form.\footnote{The same arguments may also explain the formation
  of prograde equatorial jets in Scott and Polvani's
  \citeyearpar{Scott08} simulations with an essentially frictionless
  shallow-water model.}

The simulations with different drag formulations show that better fits
to observations can be obtained if different drag formulations are
used for the different giant planets. This is physically justifiable
because the interior properties of the planets differ and give rise to
differences in the strength of MHD drag and in the depth at which it
acts \citep{Liu08}.

\section{Mean meridional circulations and angular momentum balance in
  deep atmospheres}
\label{s:AM_balance}

Our theory and simulations are consistent with the energy and angular
momentum balances of the giant planets as far as they are known, and
they are broadly consistent with many observed upper-atmospheric flow
features. Their relevance, however, depends on how the flows in the
upper atmospheres couple to flows at depth. We have represented this
coupling in an idealized fashion in our thin-shell simulations through
the drag formulation. Here we show how results for the
upper-tropospheric flows constrain the flows at depth. What follows is
a straightforward generalization of well known results for thin
atmospheres---particularly the principle of ``downward control''
\citep{Haynes91}---which was already sketched in SL09. We give the
arguments in some detail, as their implications for planetary
atmospheres are underappreciated.

\subsection{Local angular momentum balance}

In any atmosphere, regardless of its constitutional law, the balance
of angular momentum around the planet's spin axis can be written as
\begin{equation}\label{e:am_balance}
 \partial_t (\rho M)
  + \div ( \rho \mathbf{u} M )
  = -\partial_\lambda p + r_\perp \rho \mathcal{D} ,
\end{equation}
where $M = M_\Omega + M_u$ is the angular momentum per unit mass,
composed of the planetary angular momentum $M_\Omega = \Omega
r_\perp^2$ and the relative angular momentum $M_u = u r_\perp$. Here,
$r_\perp = r \cos \phi$ is the (cylindrically radial) distance to the
planet's spin axis and $r$ the (radial) distance to the planet's
center; $\mathcal{D}$ is a zonal drag force per unit mass, which may
include viscous dissipation; $\lambda$ is longitude (azimuth); and $p$
is pressure and $\rho$ density
\citep[e.g.,][chapter~11]{Peixoto92}. In the thin-shell approximation,
the distance to the spin axis is approximated as $r_\perp = a
\cos\phi$, with a constant planetary radius $a$. But the angular
momentum balance \eqref{e:am_balance} also holds in a deep atmosphere
if $r_\perp = r\cos\phi$ is taken to be the actual distance to the
spin axis, with variable $r$.

In a statistically steady state, upon averaging temporally and zonally
(azimuthally), the angular momentum balance becomes
\begin{equation}\label{e:am_balance_avg}
 \massavg{\mathbf{u}} \cdot \grad  M_\Omega
  + \massavg{\mathbf{u}} \cdot \grad \massavg{M}_u
  = r_\perp \massavg{\mathcal{D}} - \mathcal{S} ,
\end{equation}
where
\begin{equation}\label{e:amfd}
\mathcal{S} = \frac {1}{\Bar\rho} \, \div \bigl( \Bar\rho \,
  \massavg{\mathbf{u}' M_u'} \bigr)
\end{equation}
is the eddy angular momentum flux divergence. The overbar
$\overline{(\cdot)}$ now denotes the temporal and zonal mean at
constant $r_\perp$, and $\massavg{(\cdot)}=\overline{(\rho \,
  \cdot)}/\Bar\rho$ denotes the corresponding density-weighted mean;
primes $(\cdot)' = (\cdot) - \massavg{(\cdot)}$ denote deviations from
the latter. The eddy angular momentum flux divergence in
Fig.~\ref{fig_streamfunction} is the pressure-coordinate analog of the
meridional component of the flux divergence \eqref{e:amfd}.

The ratio of the second to the first term on the left-hand side of the
angular momentum balance \eqref{e:am_balance_avg} is of order Rossby
number,
\begin{equation}
  \mathrm{Ro} = \frac{U}{2\Omega L_\perp},
\end{equation}
where $L_\perp$ is the length scale of flow variations in the
cylindrically radial direction. That is, if $L$ is a meridional length
scale, $L_\perp = L \sin\phi$ is the projection of the meridional
length scale onto the equatorial plane, and $\mathrm{Ro}=U/|fL|$
becomes the familiar Rossby number for the thin-shell
approximation. Away from the equator, the Rossby number is generally
small in the tropospheres of the giant planets if zonal flow
velocities at depth do not substantially exceed those observed on
Jupiter and Saturn, or those seen in the tropospheres of Uranus and
Neptune in our simulations (see SL09 and footnote~\ref{fn:Ro}).  The
angular momentum balance then is approximately
\begin{equation}\label{e:am_balance_app}
 \massavg{\mathbf{u}} \cdot \grad  M_\Omega
  \approx r_\perp \massavg{\mathcal{D}} - \mathcal{S}.
\end{equation}
The term on the left-hand side represents the advection of planetary
angular momentum by the mean flow, or the Coriolis torque per unit
mass ($\massavg{\mathbf{u}} \cdot \grad M_\Omega = - f \, \massavg{v}
\, r_\perp$ in the thin-shell approximation). Three special dominant
balances can be distinguished.

\subsubsection{$\mathcal{D} \approx 0$, $\mathcal{S} \ne 0$}

This is the dominant balance in the off-equatorial upper troposphere,
where eddy angular momentum flux divergences are significant but drag
forces are negligible. In this case, the angular momentum balance
\begin{equation}
  \massavg{\mathbf{u}} \cdot \grad  M_\Omega \approx - \mathcal{S}
\end{equation}
implies that the mean mass flux has a component across $M_\Omega$
surfaces: toward the planet's spin axis (poleward) where eddy angular
momentum fluxes diverge ($\mathcal{S} > 0$), and away from the spin
axis (equatorward) where they converge ($\mathcal{S} < 0$). Because
eddy angular momentum fluxes in our simulations generally diverge in
retrograde tropospheric jets, or zonal velocity minima, and converge
in prograde tropospheric jets, or zonal velocity maxima, the mean
meridional mass flux in the off-equatorial upper troposphere is
generally poleward in retrograde jets and equatorward in prograde jets
(Fig.~\ref{fig_streamfunction}). The same is almost certainly true on
Jupiter and Saturn, where similar eddy angular momentum fluxes in the
upper troposphere have been observed \citep{Salyk06,DelGenio07}. (Eddy
angular momentum fluxes have not been observed on the other giant
planets.)  Direct observations of mean meridional mass fluxes on
Jupiter and Saturn are ambiguous, but the observed distribution of
convection provides indirect evidence that there is upwelling in the
cyclonic shear zones between retrograde and prograde jets
\citep{Ingersoll00,Porco03,DelGenio07}, consistent with these
arguments and with our simulations (Fig.~\ref{fig_streamfunction}).

\subsubsection{$\mathcal{D} \approx 0$, $\mathcal{S} \approx 0$}

This is the dominant off-equatorial balance immediately below the
layer with significant eddy angular momentum flux divergences, where
drag forces are negligible. In this case, the angular momentum balance
\begin{equation}\label{e:along_Mo}
  \massavg{\mathbf{u}} \cdot \grad  M_\Omega \approx 0
\end{equation}
implies that the mean mass flux is along $M_\Omega$ surfaces, that is,
parallel to the planet's spin axis in deep atmospheres or vertical in
thin atmospheres. As in Earth's atmosphere, such an off-equatorial
tropospheric layer with mean mass flux along $M_\Omega$ surfaces is
clearly seen in our simulations, where $M_\Omega$ surfaces are
vertical (Fig.~\ref{fig_streamfunction}). It very likely also exists
at least on Jupiter and Saturn, where, as we argued in the
introduction, energetic constraints indicate that significant eddy
angular momentum fluxes cannot extend deeply into the atmosphere.

The constraint that the mean mass flux is along $M_\Omega$ surfaces is
not to be confused with the Taylor-Proudman constraint. The
Taylor-Proudman constraint states that steady-state velocities in
rapidly rotating barotropic atmospheres do not vary in the direction
of the planet's spin axis if non-conservative forces are absent
\citep[e.g.,][]{Kaspi09}. It requires the flow to be barotropic,
whereas the flows we consider generally are baroclinic and sheared
along $M_\Omega$ surfaces, as in Earth's atmosphere and in our
simulations (e.g., Figs.~\ref{fig_circulation} and
\ref{fig:pottemp_temp}).

\subsubsection{$\mathcal{D} \ne 0$, $\mathcal{S} \approx 0$}

This is the dominant off-equatorial balance (Ekman balance) in lower
layers in our simulations, where drag forces are significant. It very
likely also is the dominant balance in any deep layer of significant
drag on the giant planets. In this case, the angular momentum balance
\begin{equation}
  \massavg{\mathbf{u}} \cdot \grad  M_\Omega \approx r_\perp
 \massavg{\mathcal{D}}
\end{equation}
implies that the mean mass flux has a component across $M_\Omega$
surfaces: toward the planet's spin axis (poleward) where the drag
force is retrograde ($\mathcal{D} < 0$), and away from the spin axis
(equatorward) where it is prograde ($\mathcal{D} > 0$). To the extent
that the drag force locally retards the mean zonal flow, as it does
for the linear drag in our simulations, it implies that away from the
equator, there is a mean mass flux toward the spin axis where the mean
zonal flow is prograde and away from the spin axis where it is
retrograde (Fig.~\ref{fig_streamfunction}).

\subsection{Mean meridional circulation and zonal flow at depth}

Thus overturning mass circulations in the meridional plane come
about. In a statistically steady state, any mean mass flux across an
$M_\Omega$ surface associated with eddy angular momentum flux
divergences in the upper troposphere must be balanced by an equal and
opposite mean mass flux across the same $M_\Omega$ surface somewhere
else, to obtain closed circulation cells. Where the Rossby number is
small, this opposing mean mass flux must be associated with an
opposing eddy angular momentum flux divergence or drag. Outside the
equatorial no-drag region in our simulations, the opposing mean mass
flux is associated with drag at depth, similar to how mass circulation
cells close in Earth's atmosphere. On the giant planets, MHD drag acts
at great depth and can fulfill a similar role in closing circulation
cells.

\begin{figure}[!htb]
\centering \includegraphics{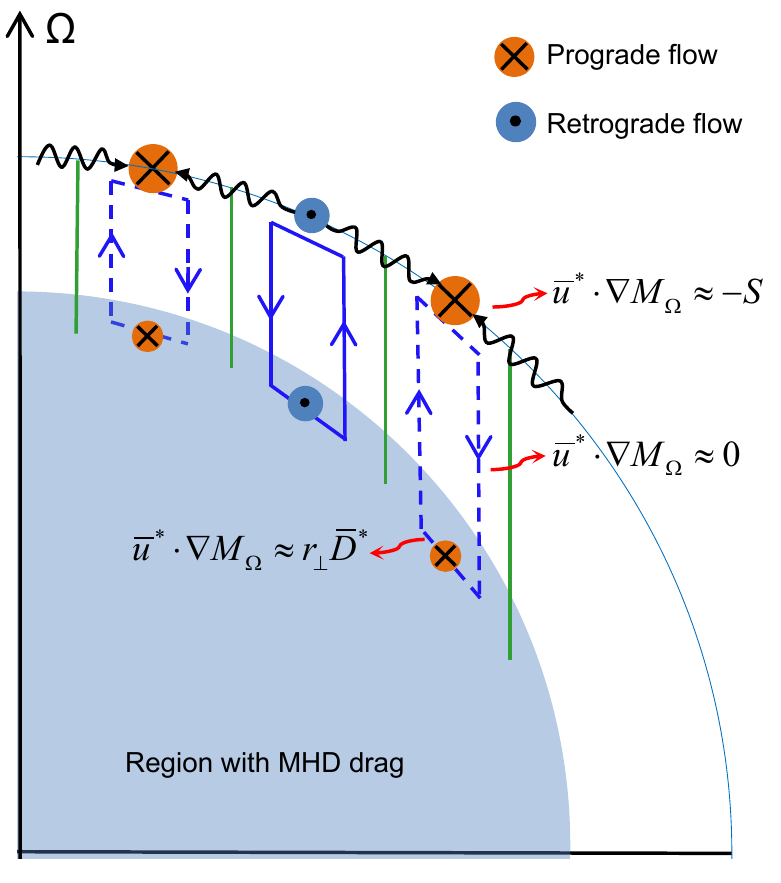}
\caption{Sketch of mean meridional circulation and zonal flow off the
  equator in giant planet atmospheres. Straight blue lines with arrows
  indicate the mass circulation; green lines indicate $M_\Omega$
  contours; wavy lines indicate eddy angular momentum fluxes. The size
  of the zonal-flow symbols is to suggest the speed of the flow. The
  blue shaded region represents the electrically conducting part of
  the atmosphere, where MHD drag acts.}
\label{fig:cartoon}
\end{figure}

The angular momentum balance also constrains the zonal flow at
depth. Taking a density-weighted integral of the angular momentum
balance \eqref{e:am_balance_app} along $M_\Omega$ surfaces and using
mass conservation shows that any net divergence or convergence of eddy
angular momentum fluxes on an $M_\Omega$ surface must be balanced by a
zonal drag force on the same $M_\Omega$ surface,
\begin{equation}\label{e:int_balance}
  \{\Bar\rho \, \mathcal{S}\}_\Omega \approx r_\perp \{ \Bar\rho \,
 \massavg{\mathcal{D}} \}_\Omega,
\end{equation}
where $\{ \cdot \}_\Omega$ denotes an average over $M_\Omega$
surfaces.  To the extent that the drag force locally retards the mean
zonal flow, it follows that if eddy angular momentum flux convergence
occurs in the upper troposphere in prograde jets, and divergence in
retrograde jets, and if this is the dominant eddy angular momentum
flux convergence/divergence on an $M_\Omega$ surface, the mean zonal
flow where the drag acts must be of the same sign as the flow in the
upper troposphere on the same $M_\Omega$ surface. That is, zonal jets
must extend to wherever drag acts, irrespective of its depth, even if
the eddy angular momentum fluxes are confined to the upper troposphere
[see \citet{OGorman08a} for a numerical example]. Because drag cannot
act at the upper boundary of the atmosphere (it would imply an
impossible torque on outer space), the jets generally extend
downward. Insofar as the eddy angular momentum fluxes in the upper
atmosphere control the dissipation at depth, one may speak of
``downward control'' of the mean meridional circulation and zonal flow
at depth \citep{Haynes91}.  Figure~\ref{fig:cartoon} summarizes these
inferences from the off-equatorial angular momentum balance.

The mean meridional circulation cells link the dynamics in the upper
troposphere to the flow at depth, adjusting the thermal structure of
the atmosphere between the upper troposphere and the layer where the
drag acts such that the zonal-flow shear along $M_\Omega$ surfaces
implied by thermal-wind balance becomes consistent with the balance
\eqref{e:int_balance} between eddy angular momentum flux divergences
and drag. This is analogous to the role mean meridional circulations
play in Earth's atmosphere \citep[e.g.,][chapter~10]{Holton04}.

Where the Rossby number is not small, the advection of angular
momentum by the mean mass flux---the second term on the left-hand side
of \eqref{e:am_balance_avg}---cannot be neglected and also contributes
to the angular momentum balance and its density-weighted integral
along $M_\Omega$ surfaces, as discussed in SL09. However, in the
tropospheres of the giant planets, this appears to be significant only
within a few degrees latitude of the equator (see SL09 and
footnote~\ref{fn:Ro}).

\subsection{Implications for thermal structure}\label{s:temperature}

Jupiter, Saturn, and Neptune have sufficiently strong intrinsic heat
fluxes to lead to convection in their tropospheres
\citep{Guillot04,Guillot05}, as in our simulations. The occurrence of
convection and the expected homogenization of entropy along convective
plumes further constrains the thermal structure of the tropospheres
and thus, through thermal wind balance, the zonal flow structures.

In the giant planet tropospheres, the convective Rossby number is
generally small, and viscous momentum dissipation and thermal
diffusion are thought to be negligible. Under these circumstances,
convective plumes are columns aligned with $M_\Omega$ surfaces,
because, as above, rapid rotation inhibits motion perpendicular to
$M_\Omega$ surfaces in the absence of viscous or other stresses
\citep[e.g.,][]{Busse76,Busse78,Kaspi09,Jones09}.\footnote{More
  generally, where the convective Rossby number is not necessarily
  small, convective plumes are aligned with angular momentum ($M$)
  surfaces.}  Therefore, convection tends to homogenize entropy along
columns in the direction of the planet's spin axis (in deep
atmospheres) or in the vertical (in thin atmospheres); however, it
does not constrain entropy gradients in perpendicular directions. The
forced-dissipative statistically steady state that results in the
presence of vigorous convection thus is neutral with respect to
slantwise convective instabilities, that is, convective and inertial
axisymmetric instabilities (\citealp{Emanuel83b,Thorpe89};
\citealp[][chapter~12]{Emanuel94}; \citealp{Schneider07b}).

In our Jupiter, Saturn, and Neptune simulations, such a state with
nearly convectively neutral interior tropospheres is indeed attained
outside a few degrees of the equator (Fig.~\ref{fig_circulation}). The
entropy and its meridional gradient hence do not vary in the vertical
(e.g., Fig.~\ref{fig:pottemp_temp}), and neither does, by thermal wind
balance, the shear (with respect to pressure) of the zonal flow. A
corresponding state with entropy and its meridional gradient
homogeneous in the direction of the spin axis can be expected to be
attained on the actual planets, and thermal wind balance for deep
atmospheres then similarly constrains the zonal flow. For a deep
atmosphere in the anelastic approximation (valid for small
fluctuations relative to an isentropic reference state), thermal wind
balance reads
\begin{equation}\label{e:thermal_wind}
  2\Omega \frac{\partial u}{\partial z} = -\frac{\alpha_s g}{r}
  \,\frac{\partial s}{\partial\phi},
\end{equation}
where $z$ is the cylindrical height coordinate in the direction of the
spin axis, $s$ is the specific entropy, and $\alpha_s(\rho)$ is a
thermal expansion coefficient that relates entropy fluctuations to
density fluctuations \citep{Kaspi09}.  That is, if the meridional
entropy gradient does not vary in the direction of the spin axis, the
shear of the zonal flow in the direction of the spin axis depends only
on $\alpha_s$, $g$, and $r$, all of which generally vary ($\alpha_s$
and $g$ vary primarily with $r$). In Jupiter's and Saturn's upper
tropospheres, the meridional entropy gradient and thus the zonal
thermal wind shear approximately vanish at the zeros of the zonal wind
\citep{Simon-Miller06,Read06,Read09b}, as they do in our simulations
(Fig.~\ref{fig_circulation}). To the extent that entropy deeper in the
troposphere is homogenized in the direction of the spin axis, the
thermal wind balance \eqref{e:thermal_wind} suggests that the zonal
flow shear then vanishes at all depths extending downward in the
direction of the spin axis from the upper-tropospheric zeros of the
zonal flow. So zeros of the zonal flow project downward along
cylinders concentric with the spin axis.\footnote{Where latent heat
  release in phase changes of water is dynamically important, a moist
  entropy rather than a dry entropy should be considered, and such a
  moist entropy can be expected to be homogenized in the direction of
  the spin axis.}

In the literature on the giant planets, it is often taken as axiomatic
that their interior tropospheres are rendered isentropic by
convection, resulting in zonal flows (Taylor columns) without shear in
the direction of the spin axis \citep[e.g.,][]{Vasavada05}. However,
convection in general does not homogenize entropy in directions
perpendicular to angular momentum surfaces, so an isentropic interior
cannot be assumed a priori. The flow on the giant planets, where the
Rossby number is small, must satisfy the constraints
\eqref{e:am_balance_app}--\eqref{e:int_balance} dictated by the
angular momentum balance, as well as thermal wind balance
\eqref{e:thermal_wind}. Given that significant eddy angular momentum
flux divergences in the upper tropospheres have been observed at least
on Jupiter and Saturn, it is very unlikely that the zonal drag, which
depends on the zonal flow, can balance the net eddy angular momentum
flux divergence on an $M_\Omega$ surface without any shear of the
zonal flow in the direction of the spin axis. It hence is very
unlikely that the interiors of the giant planets are
isentropic.\footnote{If entropy deviations from an isentropic
  reference state are not small so that the anelastic approximation
  cannot be made, the thermal wind balance \eqref{e:thermal_wind}
  becomes more complicated \citep{Kaspi09}. But this does not affect
  the qualitative considerations on which our conclusions are based.}

\subsection{Implications for role of drag}\label{s:AM_drag}

Significant eddy angular momentum flux divergences have been observed
in Jupiter's and Saturn's upper tropospheres, but no mechanism has
been proposed of how they could be balanced by opposing eddy angular
momentum flux divergences at depth such the angular momentum balance
\eqref{e:int_balance} integrated over $M_\Omega$ surfaces is satisfied
without a drag mechanism. On the other hand, coupling of the flow at
depths at which the atmosphere is electrically conducting to the
magnetic field is a plausible mechanism that generates MHD drag
(\citealp{Liu08}; see also the appendix). This MHD drag can close the
angular momentum balance integrated over $M_\Omega$
surfaces. Therefore, we adopted as our working hypothesis that the MHD
drag acting at depth couples to the flow in the upper atmosphere,
although details such as how the MHD drag depends on the zonal flow
are poorly understood. This approach gave statistically steady states
in which the angular momentum balance is closed in a manner that is
physically plausible and consistent with
observations.

Outside a few degrees latitude of the equator, the upper-tropospheric
dynamics are linked to the flow and any drag at depth along
cylindrical $M_\Omega$ surfaces. Therefore, outside the tangent
cylinder that just grazes the region of substantial MHD drag in the
equatorial plane, the upper-tropospheric flow cannot be linked to drag
at depth. We chose to represent this equatorial region of no effective
drag on the upper-tropospheric flow in our thin-shell simulations by
having an equatorial no-drag region. A no-drag region extending to
$33^\circ$ latitude corresponds to assuming that the region of
substantial MHD drag is confined within about $\cos (33^\circ) = 0.84$
planetary radii. It is doubtful that this is an accurate estimate for
all giant planets \citep{Liu08}. However, as we have shown in
section~\ref{s:Mechanisms}\ref{s:drag}, where exactly the MHD drag
acts, and how strong it is, does not affect the essence of our
results.

The preceding discussion shows that simulations of only thin
atmospheric shells can have closed energy and angular momentum
balances that are physically plausible and consistent with
observations of the giant planets. A model domain of sufficient depth
to take into account the absorption of solar radiation in the upper
atmosphere is essential to obtain baroclinic flows with an energy
balance that is consistent with observations. Resolving baroclinic
eddy fluxes of angular momentum in the upper atmosphere is essential
to obtain an angular momentum balance that is consistent with
observations. Drag at depth closes the balances in a physically
plausible manner (e.g., without assuming excessive viscous stresses in
the planetary interior).  However, the depth of the nearly inviscid
interior layer in which there are no significant heat sources and
where $\mathcal{D} \approx 0$ and $\mathcal{S} \approx 0$ is
immaterial for the mechanisms we discussed.\footnote{Latent heat
  release in phase changes of water may play a role in that layer, but
  it does not represent an external heat source, merely a conversion
  between forms of energy, and hence it does not affect the integrated
  energy balance.} The depth of that layer can be expected to affect
quantitative aspects such as the vertical shear of the zonal flow, but
we do not expect it to affect the qualitative aspects and large-scale
flow features on which we have focused.

\section{Conclusions}\label{s:Conclusion}

We have presented the first simulations of all four giant planets with
closed energy and angular momentum balances that are consistent with
observations. The simulations reproduce many large-scale features of
the observed flows, such as equatorial superrotation on Jupiter and
Saturn and equatorial subrotation on Uranus and Neptune. They exhibit
temperature structures that are broadly consistent with available
observations, and they reproduce many details of the observed flows,
for example, their vertical structure to the extent it is known and
characteristic equatorial waves observed on Jupiter. We have
demonstrated that equatorial superrotation is generated if convective
Rossby wave generation is strong and low-latitude angular momentum
flux divergence owing to baroclinic eddies generated off the equator
is sufficiently weak (Jupiter and Saturn); equatorial subrotation
results if either convective Rossby wave generation is weak or absent
(Uranus) or low-latitude angular momentum flux divergence owing to
baroclinic eddies is sufficiently strong (Neptune).

Current computational resources limit our ability to simulate flows at
depth.  However, considerations of the angular momentum balance have
shown that the zonal jets should extend---generally with shear---to
the depth where drag acts on them and balances the angular momentum
flux divergences and convergences in the upper troposphere.  That drag
acts on the zonal flow at depth is suggested by observations of eddy
angular momentum fluxes on Jupiter and Saturn, and a plausible MHD
drag mechanism exists. Though quantitative aspects (e.g., jet strength
and shear) may be affected by our inability to resolve the flow and
drag at depth, the jet formation mechanisms we discussed are not
affected by it.

We expect as-yet unobserved aspects of the flow and temperature
structures to be consistent with the simulations and mechanisms we
presented. For example, we predict that NASA's upcoming JUNO mission
to Jupiter will find evidence of zonal flows with vertical shear
similar to those in Fig.~\ref{fig_circulation}: near the equator, a
strong and deep prograde jet, and away from the equator, prograde jets
that weaken and retrograde jets that weaken only slightly or
strengthen with depth. The thermal and gravitational signature of such
zonal flows will likely be measurable by JUNO.

\paragraph{Acknowledgments}
This work was supported by a David and Lucile Packard Fellowship. The
GCM is based on the Flexible Modeling System of the Geophysical Fluid
Dynamics Laboratory; the simulations were performed on Caltech's
Division of Geological and Planetary Sciences Dell cluster. We thank
Andy Ingersoll and Yohai Kaspi for helpful discussions and comments on
drafts of this paper.


\appendix{A}
\centerline{\textbf{General Circulation Model}}
\medskip\nopagebreak%

\subsection{Resolution}

The GCM solves the hydrostatic primitive equations using the spectral
transform method in the horizontal and finite differences in the
vertical. The horizontal spectral resolution depends on the radius of
the planet being simulated (Table~1). The vertical coordinate is
$\sigma = p/p_s$ (pressure $p$ normalized by pressure at lower
boundary $p_s$); it is discretized with 30 levels for Jupiter and
Saturn and 40 levels for Uranus and Neptune.

\subsection{Drag at lower boundary}

All parameter choices are constrained by knowledge of the physical
properties of the planets and material properties of their
atmospheres, as well as by observations where available. However, the
drag formulation at the artificial lower boundary of the GCM is poorly
constrained by data or physics. It represents the MHD drag the flow
experiences in the interior of the planets.

In the interior of Jupiter and Saturn, the conductivity of molecular
hydrogen increases with depth and becomes approximately constant where
hydrogen becomes metallic at $\about 1.4 \, \mathrm{Mbar}$
\citep{Nellis96}. In the interior of Uranus and Neptune, the
conductivity of the gas envelope likewise increases with depth and is
determined by the conductivity of hydrogen and water ice
\citep{Nellis97}. In the high-conductivity interior, the interaction
of the magnetic field and the fluid flow produces Ohmic dissipation
and retards the flow \citep{Liu08}.

We represented this MHD drag deep in the atmosphere in the simplest
possible way in our thin-shell GCM, choosing the same drag formulation
and depth of the artificial lower boundary in all giant planet
simulations, to rule out that differences among them are caused by
differences in poorly constrained parameters. We assume linear
(Rayleigh) drag acts near the GCM's lower boundary, but only outside
an equatorial latitude band (see SL09 and
section~\ref{s:AM_balance}\ref{s:AM_drag}). As in the models in SL09
or \citet{Held94}, the drag coefficient decreases linearly from its
value $k_0$ at the lower boundary at $\sigma = 1.0$ to zero above
$\sigma=0.8$. The equatorial no-drag region extends to
$\phi_0=33^\circ$ latitude in all our simulations, corresponding to a
MHD drag that acts only within $0.84$ planetary radii. The drag
coefficient is constant ($k_0=10^{-2}\,\mathrm{days^{-1}}$) outside
this region. The kinetic energy dissipated by the Rayleigh drag (a few
percent of the sum of the intrinsic heat flux and the absorbed solar
radiative flux) is returned to the flow locally as heat to conserve
energy.

We chose the width of the no-drag region and the drag coefficient
outside of it empirically, to obtain jets in the upper atmosphere that
have similar strength and width as the observed jets. By choosing drag
formulations that differ from planet to planet, better fits to
observations could be obtained (cf.\
section~\ref{s:Mechanisms}\ref{s:drag}).

\subsection{Radiative transfer}

As in SL09, radiative transfer is represented as that in a homogeneous
gray atmosphere, using the two-stream approximation. The
top-of-atmosphere (TOA) insolation is imposed as perpetual equinox
with no diurnal cycle, $F_\mathrm{TOA} = (F_0/\pi) \cos\phi$, with the
appropriate solar constant $F_0$ for each planet (Table~1). That is,
for the purposes of this paper, we ignore the seasonal cycle in TOA
insolation, which is substantial for Saturn, Uranus, and Neptune
because of their obliquities. Ignoring seasonality may be justifiable,
for example, if response timescales of the atmospheric circulation
(e.g., radiative timescales) are much longer than seasonal timescales,
which may be the case on the giant planets. However, the nonzero
obliquities also influence the annual-mean TOA insolation, especially
on Uranus, an influence we likewise ignore.

The solar radiative flux for a semi-infinite scattering and absorbing
atmosphere is calculated for a solar optical depth $\tau_s$ that is
linear in pressure, $\tau_s = \tau_{s0} (p/p_0)$, to represent
scattering and absorption by a well-mixed absorber. The solar optical
properties of the giant planet atmospheres are not well
constrained. To minimize differences among the simulations, we chose
the same solar optical properties for all giant planets: $\tau_{s0} =
3.0$ at $p_0 = 3.0$~bar. This gives a solar radiative flux
qualitatively consistent with \emph{Galileo} probe measurements in
Jupiter \citep{Sromovsky98}.

The thermal radiative flux is calculated for a gray atmosphere in
which the thermal optical depth $\tau_l$ is quadratic in pressure,
$\tau_l = \tau_{l0} (p/p_0)^2$, to represent collision-induced
absorption of thermal radiation. The thermal optical depths
$\tau_{l0}$ at pressure $p_0$ are chosen such that the observed
thermal emission levels \citep[e.g.,][]{Ingersoll90} of the giant
planets approximately correspond to $\tau_l = 1$.  The thermal optical
depths thus vary from planet to planet (Table~1).

The lower boundary condition for the radiative fluxes is energy
conservation: the upward thermal radiative flux is set equal to the
sum of the downward solar flux and thermal radiative flux at each grid
point.

\subsection{Intrinsic heat flux}

A spatially uniform and temporally constant heat flux, corresponding
to that estimated for the giant planets (Table~1), is
deposited in the GCM's lowest layer to mimic intrinsic heat fluxes.

\subsection{Convection scheme}

A quasi-equilibrium convection scheme represents (dry) convection. It
relaxes temperature profiles toward a convective profile with
adiabatic lapse rate $\Gamma = g/c_p$ \citep{Schneider06b}.  The
convective relaxation time is chosen to be roughly the time it takes a
gravity wave with speed $c$ to traverse the extratropical Rossby
radius $L_x = c/f$, that is, roughly an extratropical inertial time
$f^{-1}$. We chose the convective relaxation time to be $6\,
\mathrm{hr}$ for Jupiter and Saturn and $10\,\mathrm{hr}$ for Uranus
and Neptune. We experimented with convective relaxation times up to a
factor 2 smaller and a factor 4 larger in preliminary simulations; the
simulated flows appeared not to be sensitive to such variations of the
relaxation time.

The convection scheme does not transport momentum, that is, it assumes
a convective Prandtl number of zero. Prandtl numbers for dry
convection are usually greater than zero, so this represents an
idealization, which may affect our results. However, the convective
generation of Rossby waves on which we focus occurs in the equatorial
upper troposphere, where the vertical shear is relatively small
(Fig.~\ref{fig_circulation}). So one may expect convective momentum
fluxes not to alter our results substantially. But to the extent that
the vertical shear and convective momentum fluxes cannot be neglected,
they may amplify the superrotation in the upper troposphere in the
Jupiter and Saturn simulations, as the equatorial zonal flow in those
simulations is stronger at depth than in the upper troposphere, so
that convection can be expected to transport momentum upward. The
dependence of our results on the convective Prandtl number is worth
investigating further.

\subsection{Subgrid-scale dissipation}

For $\sigma \le 0.8$, above the layer with Rayleigh drag, horizontal
hyperdiffusion in the vorticity, divergence, and temperature equations
is the only frictional process. The hyperdiffusion is represented by
an exponential cutoff filter \citep{Smith02b}, with a damping time scale
of 2~h at the smallest resolved scale. The cutoff wavenumber depends
on the horizontal resolution (Table~1).

The energy dissipated by the subgrid-scale hyperdiffusion is not
returned to the flow as heat. However, it amounts to less than $1\%$
of the total energy uptake of the atmosphere in all simulations.

\subsection{Simulations}

The simulations were spun-up from an isothermal rest state, with small
perturbations in temperature and vorticity to break the axisymmetry of
the initial state. Each simulation was integrated for at least
40,000~Earth days. In the statistically steady states, the global-mean
outgoing thermal radiative flux is within $1 \%$ of the sum of the
global-mean solar radiative flux and the imposed intrinsic heat
flux. The vertically integrated Rayleigh drag on the zonal flow
approximately balances the vertically integrated total (mean plus
eddy) angular momentum flux convergence. The circulation statistics
shown are computed from flow fields sampled 4 times daily and averaged
over 1500~days. They were first computed on the GCM's sigma surfaces,
with the appropriate surface pressure-weighting of the averages
\citep{Schneider06b}, and then interpolated to pressure surfaces for
display purposes.


\end{document}